\documentclass[12pt]{article}
\usepackage{graphicx}
\usepackage{amsmath}
\newcommand{\sm}{\hbox{$\bigcirc$\kern-0.72em\hbox{\bf s} }}
\newcommand{\Id}{\hbox{\sl 1\kern-0.25em\hbox{I}}}

\newcommand{\rcorr}{\hbox{\kern-1.2em$\longrightarrow$}}
\newcommand{\lrcorr}{\hbox{\kern-1.2em$\longleftrightarrow$}}

\newcommand{\nRightarrow}{\Rightarrow\kern-1.2em\hbox{/}\kern.8em} %
\newcommand{\BB}{\hbox{I\kern-.2em\hbox{B}}} 
\newcommand{\DD}{\hbox{I\kern-.2em\hbox{D}}} 
\newcommand{\FF}{\hbox{I\kern-.2em\hbox{F}}} 
\newcommand{\NN}{\hbox{I\kern-.2em\hbox{N}}}  
\newcommand{\ZZ}{{{\rm Z}\kern-.28em{\rm Z}}} 
\newcommand{\RR}{\mathop{{\rm I}\kern-.2em{\rm R}}\nolimits} 
\newcommand{\RRe}{\mathop{{\rm I}\kern-.2em{\rm Re}}\nolimits} 
\newcommand{\QQ}{\hbox{l\kern-.36em\hbox{Q}}}  
\newcommand{\CC}{\hbox{{\textsf I}\kern-.47em\hbox{C}}}
\newcommand{\nop}{\hbox{{\textsf I}\kern-.47em\hbox{O}}}

\def\TREV{{{}^\triangleleft\kern-1.5pt\texttt{T}}}
\def\trev{{{}^\triangleleft\kern-3.2pt\texttt{t}}}
\def\SREV{{{}_\triangleleft\kern-2pt\texttt{S}}}
\def\srev{{{}_\triangleleft\kern-2.2pt\texttt{s}}}
\begin{document}
\title{Relativistic Particle Theories\\
Without Canonical Quantization}
\author{Giuseppe Nistic\`o
\\
{\small Dipartimento
di Matematica e Informatica, Universit\`a della Calabria, Italy}\\
{\small and}\\
{\small
INFN -- gruppo collegato di Cosenza, Italy}\\
{\small email: giuseppe.nistico@unical.it} } \maketitle
\abstract{The difficulties of relativistic particle theories formulated my means of canonical quantization,
such as Klein-Gordon and Dirac theories, ultimately led theoretical physicists to turn on quantum field theory
to model elementary particle physics. The aim of the present work is to pursue a method alternative
to canonical quantization that avoids these difficulties. In order to guarantee this result,
the present approach is constrained to be developed {\sl deductively from physical principles}.
The physical principles assumed for a free particle consist of
the symmetry properties of the particle with respect to the Poincar\'e group and of the transformation properties
of the position observable, expressed by means of a suitably conceived notion of {\sl quantum transformation}.
In so doing, the effectiveness of group theoretical methods is exploited.
\par
Our work has pointed out the necessity of new classes of irreducible representations of the Poincar\'e group the theory can be based on.
For spin 0 particle, four inequivalent theories are completely determined, that differ from Klein-Gordon theory
for the physical interpretation but also for the mathematical structures. In particular, two of these theories requires
irreducible representations in the new classes.
\par
The present lack of the explicit transformation properties of position with respect to boosts forbids the complete
determination of {\sl non zero} spin particle theories.
A particular form of these transformation
properties was in past adopted by Jordan and Mukunda. We check its consistency within the present approach; it is
found that for spin $\frac{1}{2}$ particles there is only one consistent theory; it is unitarily related to Dirac theory, but, once again,
it requires the new classes of irreducible representations.
}
\section{Introduction}
\subsection{Motivation and methodological commitments}
The explicit formulation of the quantum theory of a {\sl specific physical system} requires
to accomplish three tasks.
\begin{itemize}
\item[{\bf T.1.}]
To identify the {\sl concrete} Hilbert space $\mathcal H$ the theory will be based on.
\item[{\bf T.2.}]
For each observable $\mathcal A$ specifying the physical system, to explicitly identify the self-adjoint operator $A$ of $\mathcal H$
that represents $\mathcal A$.
\item[{\bf T.3.}]
To explicitly identify the Hamiltonian operator that rules
over time evolution.
\end{itemize}
{\sl Canonical quantization} was the primary method for formulating {\sl specific
relativistic particle theories} \cite{Klein},\cite{Fock},\cite{Gordon};
differently from the non-relativistic case, the results were affected
by problems. For instance,
the wave equation for a spin-0 free particle, i.e. Klein-Gordon equation
$$
\left\{\frac{\partial^2}{\partial t^2}-\left(\frac{\partial^2}{\partial x_1^2}+
\frac{\partial^2}{\partial x_2^2}+\frac{\partial^2}{\partial x_3^2}\right)\right\}\psi_t({\bf x})=
-m^2\psi_t({\bf x}),\eqno{\hbox{(K-G)}}
$$
turned out to be {\sl second order} with respect to time, while according to the general law of quantum theory it should be {\sl first order}.
Another problem is the inconsistency of the four-density concept viable in the theory \cite{BM68}.
\par
Since equation (K-G)
does not admit ground states, i.e. there are arbitrarily large negative energy levels,
further problems arose when interaction is taken into account;
Dirac argued \cite{Dir28} that
interaction with radiation would cause transitions of electrons into the lower negative levels.
Positive energy electrons would be consequently ``swallowed'' by the infinite negative levels,
in contrast with the stability of matter.
To solve the problems
Dirac \cite{Dir28} proposed his equation for electrons,
$$
i\frac{d}{dt}\psi_t=(2\rho_1{\bf S}\cdot{\bf p}+m\rho_3)\psi_t,
$$
that is first order; yet, ground states do not exist.
In order that the theory made sense, Dirac \cite{Dir30} proposed the
further hypothesis that almost all negative levels
are occupied, so that transitions of electrons into a negative energy state is forbidden by Pauli exclusion principle, and the few
holes in this negative energy {\sl sea} are the {\sl protons}. The inconsistency of this hypothesis with the complete symmetry between positive and negative energy states proved by Weyl \cite{Weyl} led Dirac \cite{Dir31} to replace protons with {\sl positrons}, the anti-particles of electrons.
However, since the effectiveness of Dirac's
hypotheses is based on the Pauli exclusion principle, his proposal does not explain the existence of
anti-particle of integral spin particles.
We see that each attempt to repair a problem generates other problems, as Weinberg more effectively showed in more detail \cite{WeinBook}.
Quoting, ``... it became generally clear that relativistic wave mechanics, in the sense of a relativistic quantum theory of
a fixed number of particles, is an impossibility.
Thus, despite its many successes, relativistic wave mechanics was ultimately to give way to quantum field theory'' \cite{WeinBook}.
\par
This state of affairs
has its roots in the methodological features of canonical quantization. In order to formulate
the quantum theory of a specific system, canonical quantization prescribes to start from its {\sl classical theory}; e.g., for a particle
we could start
from its Hamiltonian classical theory, where the
position's coordinates $(q_1,q_2,q_3)\equiv{\bf q}$ are dynamical variables, with
conjugate momenta ${\bf p}\equiv(p_1,p_2,p_3)$ and with Hamiltonian function $h({\bf q},{\bf p},t)$;
then the prescriptions of canonical quantization dictat to replace the dynamical variables $q_j$ and their momenta $p_j$ with
operators $Q_j$ and $P_j$, and to replace the Poisson brackets $\{\;,\;\}$ with operator's commutators $i[\;,\;]$
in the equations of the classical theory, so that the time evolution equations turn out to be
$$
\frac{d}{dt}Q_j^{(t)}=i[h({\bf Q},{\bf P},t), Q_j^{(t)}]\;,\;\;\;\frac{d}{dt}P_j^{(t)}=i[h({\bf Q},{\bf P},t), P_j^{(t)}]\;.
$$
We see that this procedure provides no deductive path that leads to results from physical principles.
For this reason the real causes of problematic or inconsistent
predictions cannot be singled out to be remedied, in general.
\par
The aim of this work is not to devise hypotheses that remedy the problems in a more effective way. Our strategy, instead,
is to pursue an approach {\sl alternative} to canonical quantization, that addresses
tasks {\bf T.1}, {\bf T.2}, {\bf T.3}
through a {\sl deductive development from physical principles}.
This strategy should prevent from the shortcomings yielded with canonical quantization,
because the eventual occurrence of inconsistencies would be the proof that some physical principle the theory is based on fails, and therefore the system of physical principles should be reconsidered to identify the valid ones.
Nevertheless, in order to maintain coherence and this epistemological quality, no assumption can be made which cannot be derived from basic principles.
These methodological commitments were effective in developing the {\sl non-relativistic} quantum theory of an interacting particle \cite{JMLL1},\cite{N1}, \cite{N2}. Here we undertake the approach for the relativistic quantum theory of an isolated system and of a free particle in particular.
\par
In fact, for an isolated system we can establish as a physical principle that {\sl Poincar\'e group $\mathcal P$ is a group of symmetry transformations}.
Then, Wigner theorem on the {\sl representation of symmetries} \cite{b1},\cite{b2} can be applied, and in so doing a first result
is implied: the Hilbert space of the theory must admit
a {\sl generalized projective representation} of $\mathcal P$, which realizes the Poincar\'e quantum transformations
of the quantum observables (self-adjoint operators) of the system just according to the well known Wigner relation \cite{b1},\cite{b2}.
\par
Therefore, the first step of the alternative approach should be to identify all irreducible generalized projective representations of $\mathcal P$.
The literature where this strategy is adopted in some extent, from the fundamental work of Wigner \cite{W2} to more recent contributions \cite{WeinBook},\cite{Costa},\cite{BarBook},\cite{JM},
does not provide a satisfactory answer to this demand. E.g., in \cite{JM} time reversal and space inversion are ignored. In \cite{c13} irreducible representations with anti-unitary space inversion are not taken into account; one reason put forward for this exclusion \cite{WeinBook},\cite{Costa}
is that anti-unitary space inversion operators entail negative spectral value of the self-adjoint generator of time translation, that in these studies is identified with {\sl energy}.
But Dirac theory of spin-$\frac{1}{2}$ particles {\sl does admit} negative values for this spectrum, so that the reasons and the conditions
for the exclusion remain not sharply determined.
On the other hand, as it is shown in the paper, the unitary or anti-unitary character of time reversal and space inversion
operators are decisive in determining the mathematical structure of the resulting theory.
There are sufficient reasons, hence, that indicate the necessity of settling the subject.
In the present work
\begin{itemize}
\item[I.]
first a self-contained re-determination is undertaken, of the possible irreducible generalized projective representations of the Poincar\'e group,
without {\it a priori} preclusions about the unitary or anti-unitary character of the time reversal and space inversion operators;
as a result two further classes of irreducible representations are found;
\item[II.]
the specific constraints implied by the peculiar features of a specific physical system are identified
in the case of a {\sl free localizable particle}, for which possible theories are derived;
in particular, it is shown not only that these constraints do {\sl not rule out} anti-unitary space inversion operators,
but also that the quantum theory of {\sl Klein-Gordon} or {\sl Dirac kind} particles can be coherently formulated only by means of the
new classes of representations.
\end{itemize}
\subsection{Summary}
Sections 2, 3 and 4 are devoted to re-determine the possible classes of irreducible generalized projective representations
of the Poincar\'e group $\mathcal P$. These sections are rather technical. Once acquainted with the notation, the reader interested in the physical implications
can skip to section 4.3 where the re-determination is resumed.
\par
In section 2 we introduce the notation and recall results of interest about relativistic groups and their representations. Then
we specialize to generalized projective representations that are irreducible.
In so doing no {\it a priori} preclusion is operated about the unitary or anti-unitary character of the time reversal and the space inversion operators.
We show how this character determines the symmetrical or asymmetrical character of the spectrum $\sigma(\underline P)$
of the self-adjoint generators $\underline P$ of translations.
Furthermore, we show that
the restriction $U\mid_{{\mathcal P}_+^\uparrow}$ of an irreducible representation of $\mathcal P$ to the proper orthochronus group ${\mathcal P}_+^\uparrow$
always admits a decomposition $U\mid_{{\mathcal P}_+^\uparrow}=U^+\mid_{{\mathcal P}_+^\uparrow}\oplus U^-\mid_{{\mathcal P}_+^\uparrow}$. Thus the first classification
(symmetrical or asymmetrical $\sigma(\underline P)$)
can be refined according to the irreducibility or reducibility of the components $U^+\mid_{{\mathcal P}_+^\uparrow}$ or $U^-\mid_{{\mathcal P}_+^\uparrow}$.
Our re-determination is carried out in sections 3 and 4 according to these two criteria.
\par
In section 3.1 we give the well known irreducible generalized projective representations
with asymmetrical spectrum $\sigma(\underline P)$, for which the reduction $U^+\mid_{{\mathcal P}_+^\uparrow}$ or $U^-\mid_{{\mathcal P}_+^\uparrow}$
is irreducible. Section 3.2 identifies all irreducible generalized projective representations
with symmetrical spectrum $\sigma(\underline P)$ and irreducible reduction $U^+\mid_{{\mathcal P}_+^\uparrow}$.
In section 4 irreducible generalized projective representations of $\mathcal P$ with the reduction
$U^+\mid_{{\mathcal P}_+^\uparrow}$ or $U^-\mid_{{\mathcal P}_+^\uparrow}$ reducible are concretely identified in both cases of symmetrical and asymmetrical
spectrum $\sigma(\underline P)$.
The re-determination, summarized in section 4.3, has identified two classes
${\mathcal I}_{\mathcal P}({\rm ant.}\SREV)$, and ${\mathcal I}_{\mathcal P}({\rm red.})$ of irreducible generalized projective representations of $\mathcal P$ that
the literature about relativistic quantum theory of single particles has not considered, at the best of our knowledge.
\vskip.5pc
In section 5 the problem of determining the possible specific relativistic quantum theories of a {\sl single free particle} is addressed.
It is shown that the invariance of the theory of {\sl any} isolated system with respect to Poincar\'e group compels the existence of
a particular transformation of quantum observables, called {\sl quantum transformation}, in correspondence with each Poincar\'e transformation. The properties of
quantum transformations
imply that the theory must possesses an irreducible generalized projective representation $g\to U_g$ of the Poincar\'e group that realize the quantum transformations according to Wigner relation.
\par
The concept of {\sl elementary free particle} is introduced in section 5.1 as an isolated system
endowed with a unique {\sl position observable} ${\bf Q}=(Q_1,Q_2,Q_3)$.
The identification of the theories of an {\sl elementary free particle} is accordingly performed in section 5.2 and 5.3 by selecting
which of the already identified irreducible generalized projective representations of $\mathcal P$, with $U^\pm\mid_{{\mathcal P}_+^\uparrow}$ irreducible,
admit a position operator $\bf Q$.
As a result, four inequivalent theories for {\sl spin 0} particles are completely determined,
which do not suffer the shortcomings of the theories obtained by canonical quantization.
Two of these theories are characterized by {\sl symmetrical} spectrum $\sigma(\underline P)$, hence recalling Klein-Gordon particles.
The fundamental differences of the derived theories with respect to
Klein-Gordon theory are highlighted in section 5.4.
\par
In the case of {\sl non zero spin} particles, the selection of the irreducible generalized projective representations that admit a unique position observable does not
uniquely determine $\bf Q$. This indeterminacy is due to the lack of the explicit form of the transformation properties of $\bf Q$
with respect to boosts. A particular form (JM) of these properties as commutation relations was adopted by Jordan and Mukunda \cite{JM}.
In section 6 the selection is performed my making tentatively use of (JM) as
transformation properties with respect to boosts.
For particles of  spin $\frac{1}{2}$ and {\sl symmetrical} spectrum $\sigma(\underline P)$ only one theory turns out to be consistent with (JM),
which is  unitarily related to Dirac theory. Also this theory requires an
irreducible generalized projective representation of $\mathcal  P$ in the new class ${\mathcal I}_{\mathcal P}({\rm ant.}\SREV)$.
\par
For non-zero spin particles and {\sl asymmetrical} spectrum $\sigma(\underline P)$, no irreducible generalized projective representation of $\mathcal P$ exists
that admits a position operator $\bf Q$ satisfying (JM).
\par
In the conclusive section 7 the future perspective stemming form the present work are indicated.
\section{Mathematical preliminaries}
Sections 2.1 and 2.2 establish mathematical elements needed for the re-determination of the possible structures
of the irreducible representations of the Poincar\'e group. In section 2.3 a  coarse grained classification is operated according to the symmetrical or asymmetrical
character of the spectrum $\sigma(\underline P)$. This classification can be refined according to the reducibility or irreducibility of
$U^+\mid_{{\mathcal P}_+^\uparrow}$ or $U^-\mid_{{\mathcal P}_+^\uparrow}$.
\subsection{Mathematical prerequisites}
\subsubsection{Hilbert space formalism}
We shall make use of the following mathematical structures developable within the formalism of a complex and separable Hilbert space $\mathcal H$, that
are of general interest also in quantum theory.
\begin{description}
\item[\quad-]
The set $\Omega({\mathcal H})$ of all self-adjoint operators of $\mathcal H$; in a quantum theory these operators represent
{\sl quantum observables}.
\item[\quad-]
The lattice $\Pi({\mathcal H})$ of all projections operators of $\mathcal H$; in a quantum theory they represent observables with spectrum $\{0,1\}$.
\item[\quad-]
The set $\Pi_1({\mathcal H})$ of all rank one orthogonal projections of $\mathcal H$.
\item[\quad-]
The set ${\mathcal S}({\mathcal H})$ of all density operators of $\mathcal H$;
in a quantum theory these operators represent {\sl quantum states}.
\item[\quad-]
The set ${\mathcal V}({\mathcal H})$ of all unitary  or anti-unitary operators of the Hilbert space $\mathcal H$.
\item[\quad-]
The set ${\mathcal U}({\mathcal H})$ of all unitary operators of $\mathcal H$; trivially,
${\mathcal U}({\mathcal H})\subseteq{\mathcal V}({\mathcal H})$ holds.
\end{description}
\subsubsection{Representations of groups}
The following definition introduces generalized notions of group representation.
\vskip.5pc\noindent
{\bf Definition 2.1.} {\sl Let $G$ be a separable, locally compact group with identity element $e$. A correspondence
$U:G\to{\mathcal V}({\mathcal H})$, $g\to U_g$, with $U_e=\Id$, is a generalized projective representation of $G$
if the following conditions are satisfied.
\begin{description}
\item[\;{\rm i)}\;\;]
A complex function $\sigma:{G}\times{G}\to{\CC}$,
called multiplier,
exists such that $U_{g_1g_2}=\sigma(g_1,g_2)U_{g_1}U_{g_2}$; the modulus $\vert\sigma(g_1,g_2)\vert$ is always 1, of course;
\item[\;{\rm ii)}\;]
for all $\phi,\psi\in\mathcal H$, the mapping $g\to\langle U_g\phi\mid\psi\rangle$ is a Borel function in $g$.
\end{description}\noindent
Whenever $U_g$ is unitary for all $g\in G$, $U$ is called
projective representation, or $\sigma$-representation.
\par
A generalized projective representation is said to be continuous if for any fixed $\psi\in\mathcal H$
the mapping $g\to U_g\psi$ from $G$ to $\mathcal H$ is continuous with respect to $g$.
}
\vskip.5pc\noindent
In \cite{N1} we have proved that the following statement holds.
\vskip.5pc\noindent
{\bf Proposition 2.1.}
{\sl If $G$ is a connected group, then every continuous generalized projective representation of $G$ is a projective
representation, i.e. $U_g\in{\mathcal U}({\mathcal H})$, for all $g\in G$.}
\vskip.5pc
Given any vector $\underline x=(x_0,{\bf x})\in\RR^4$, we call $x_0$ the {\sl time component} of $\underline x$
and ${\bf x}=(x_1,x_2,x_3)$ the {\sl spatial component} of $\underline x$.
The Euclidean group $\mathcal E$ is the sub-group generated by all ${\mathcal T}_j$ and all ${\mathcal R}_j$.
The proper orthochronous Poincar\'e group ${\mathcal P}_+^\uparrow$ is the separable locally compact
group of all transformations of $\RR^4$
generated by the ten one-parameter sub-groups ${\mathcal T}_0$, ${\mathcal T}_j,{\mathcal R}_j$, ${\mathcal B}_j$, $j=1,2,3$,
of time translations, spatial translation, proper spatial rotations and Lorentz boosts, respectively. The sub-group generated by
${\mathcal R}_j$, ${\mathcal B}_j$ is the proper orthochronous Lorentz group ${\mathcal L}_+^\uparrow$ \cite{BarBook}.
It does not include time reversal $\trev$ and space inversion $\srev$.
Time reversal $\trev$ transforms $\underline x=(x_0,{\bf x})$ into $(-x_0,{\bf x})$; space inversion $\srev$
transforms $\underline x=(x_0,{\bf x})$ into $(x_0,-{\bf x})$.
The group generated by $\{{\mathcal P}_+^\uparrow, \trev,\srev\}$ is the separable and locally compact Poincar\'e group $\mathcal P$.
By ${\mathcal L}_+$ we denote the subgroup generated by ${\mathcal L}_+^\uparrow$ and $\trev$, while ${\mathcal L}^\uparrow$
denotes the subgroup generated by ${\mathcal L}_+$ and $\srev$; analogously, ${\mathcal P}_+$ denotes
the subgroup generated by ${\mathcal P}_+^\uparrow$ and $\trev$, while ${\mathcal P}^\uparrow$ is the subgroup
generated by ${\mathcal P}_+^\uparrow$ and $\srev$.
\vskip.5pc
Since ${\mathcal P}_+^\uparrow$ is a {\sl connected} group, according to Prop. 2.1
every continuous projective representation of ${\mathcal P}_+^\uparrow$ is a projective representation.
In sections 2,3,4 we consider only irreducible generalized projective representation of $\mathcal P$
whose restriction to ${\mathcal P}_+^\uparrow$ is continuous, so that $U_g\in{\mathcal U}({\mathcal H})$
for all $g\in{\mathcal P}_+^\uparrow$.
\par
All sub-groups ${\mathcal T}_0$, ${\mathcal T}_j,{\mathcal R}_j$, ${\mathcal B}_j$ of ${\mathcal P}_+^\uparrow$ are additive;
in fact, ${\mathcal B}_j$ is not additive with respect to the parameter
{\sl relative velocity} $u$, but it is additive with respect to the parameter $\varphi(u)=\frac{1}{2}\ln\frac{1+u}{1-u}$.
Then, according to Stone's theorem \cite{c9}, for every continuous projective representation $U$ of ${\mathcal P}_+^\uparrow$ there exist
ten self-adjoint generators $P_0$, $P_j$, $J_j$, $K_j$, $j=1,2,3$, of the ten
one-parameter unitary subgroups $\{e^{iP_0t}\}$,   $\{e^{-i{P}_j a_j},\,a\in{\RR}\}$,
$\{e^{-i{J}_j \theta_j},\,\theta_j\in{\RR}\}$, $\{e^{-i{K}_j \varphi(u_j)},\,u_j\in{\RR}\}$
that represent the one-parameter
sub-groups ${\mathcal T}_0$, ${\mathcal T}_j,{\mathcal R}_j$, ${\mathcal B}_j$ according to the projective representation
$g\to {U}_g$ of the Poincar\'e group ${\mathcal P}_+^\uparrow$.
The
structural properties of ${\mathcal P}_+^\uparrow$ as a Lie group imply that the following
commutation relations \cite{c13} hold.
\vskip.5pc\noindent
(i)\hskip4.3mm $[{P}_j,{P}_k]=\nop$,\hskip15mm (ii) $[{J}_j,{P}_k]=i{\hat\epsilon}_{jkl}{P}_l$,
\hskip8mm (iii) $[{ J}_j,{ J}_k]=i{\hat\epsilon}_{jkl}{ J}_l$,\par\noindent
(iv)\hskip2.3mm $[{ J}_j,{ K}_k]
=i{\hat\epsilon}_{jkl}{ K}_l$,\quad\hskip1.8mm (v)\hskip1mm $[{ K}_j,{ K}_k]=-i\delta_{j,k}J_l$,
\quad (vi) $[{ K}_j,{ P}_k]=i\delta_{jk}P_0$,\hfill{}(1)
\par\noindent
(vii) $[P_j,P_0]=\nop$,\hskip15.2mm (viii) $[J_j,P_0]=\nop$,\hskip14.8mm (ix) $[K_j,P_0]=iP_0$,
\vskip.4pc\noindent
where ${\hat\epsilon}_{jkl}$ is the Levi-Civita symbol ${\epsilon}_{jkl}$
restricted by the condition $j\neq l\neq k$. The following proposition holds \cite{WB}.
\vskip.5pc\noindent
{\bf Proposition 2.2.}
{\sl
Relations (1) imply
$$
[P_0^2-(P_1^2+P_2^2+P_3^2),U_g]=\nop\,,\hbox{ for all } g\in{{\mathcal P}_+^\uparrow}\,,\eqno(2)
$$
$$
[W_0^2
-(W_1^2+W_2^2+W_3^2),U_g]=\nop\,,\hbox{ for all }g\in{{\mathcal P}_+^\uparrow}\,,\eqno(3)
$$
where $W_0={\bf P}\cdot{\bf J}$ and $W_j=P_0J_j-({\bf P}\times{\bf K})_j$ form the Pauli-Lubanski four-operator
$\underline W=(W_0,{\bf W})$.}
\vskip.5pc\noindent
Accordingly, we can state the following proposition.
\vskip.4pc\noindent
{\bf Proposition 2.3.}
{\sl
Properties (2), (3) imply that if a projective representation $U:{\mathcal P}_+^\uparrow\to{\mathcal U}({\mathcal H})$
is {\sl irreducible}, then two real numbers $\eta$, $\varpi$ exist such that the following equalities hold.
$$
(i)\;\;P_0^2-{\bf P}^2=\eta\Id\quad\hbox{and}\quad(ii)\;\;W^2\equiv W_0^2
-(W_1^2+W_2^2+W_3^2)=\varpi\Id\,.\eqno(4)
$$}
\subsection{Commutation rules for $\TREV$ and $\SREV$}
Let $U:{\mathcal P}\to{\mathcal V}({\mathcal H})$ be a generalized projective representation
whose restriction to ${\mathcal P}_+^\uparrow$ is continuous. Each operator $U_g$ can be unitary or anti-unitary, but
according to Prop. 2.1 $U_g$ is unitary if $g\in{\mathcal P}_+^\uparrow$.
Since time reversal
$\trev$ and space inversion $\srev$ are {\sl not connected} with the identity transformation $e\in\mathcal P$,
the operators $\TREV=U_\trev$ and $\SREV=U_\srev$ that represent $\trev$ and $\srev$ according to Wigner theorem
are not necessarily unitary: each of them can be unitary or anti-unitary.
The commutation relations (1) must be generalized to the extended representation, by including
time reversal and space inversion operators, $\TREV$ and $\SREV$ respectively.
By making use of the structural properties of the full Poincar\'e group $\mathcal P$ \cite{c13};
the following implications are derived, that show how the relations depend on the unitary or
anti-unitary character of $\TREV$ an $\SREV$.
\vskip.5pc\noindent
If $\SREV$ {\sl is unitary}, then the following relations hold.
$$
[\SREV,P_0]=\nop,\quad \SREV, P_j=-P_j\SREV,\quad [\SREV,J_k]=\nop,\quad \SREV K_j=-K_j\SREV;\eqno(5)
$$
moreover, the free phase factor can be chosen so that $\SREV^2=\Id$, i.e. $\SREV^{-1}=\SREV$.
\vskip.4pc\noindent
In the case that $\SREV$ {\sl is anti-unitary}, instead we have
$$
\SREV P_0=-P_0\SREV,\quad [\SREV, P_j]=\nop,\quad \SREV J_k=-J_k\SREV,\quad \SREV K_j=K_j\SREV\,;\eqno(6)
$$
moreover, $\SREV^2=c\Id$, so that $\SREV^{-1}=c\SREV$, where $c=1$ or $c=-1$.
\vskip.5pc\noindent
The following relations hold in the case that $\TREV$ {\sl is unitary}.
$$
\TREV P_0=-P_0\TREV,\quad [\TREV,P_j]=\nop,\quad [\TREV,J_k]=\nop,\quad \TREV K_j=-K_j\TREV;\eqno(7)
$$
the free phase factor can be chosen so that $\TREV^2=\Id$, i.e. $\TREV^{-1}=\TREV$.
\vskip.4pc\noindent
Instead, for {\sl anti-unitary} $\TREV$ we have
$$
\TREV P_0=P_0\TREV,\quad \TREV P_j=-P_j\TREV,\quad \TREV J_k=-J_k\TREV,\quad \TREV K_j=K_j\TREV.\eqno(8)
$$
moreover, $\TREV^2=c\Id$, so that $\TREV^{-1}=c\TREV$, either $c=1$ or $c=-1$ must hold.
\vskip.5pc\noindent
The condition to be satisfied by commutator $[\SREV,\TREV]$, independently of their unitary or anti-unitary character
of $\SREV$ and $\TREV$, is simply
$$
\SREV\TREV=\omega\TREV\SREV,\hbox{ with }\omega\in\CC\hbox{ and } \vert\omega\vert=1.\eqno(9)
$$
The further relations (5)-(8) allow to extend Prop. 2.2.
\vskip.5pc\noindent
{\bf Proposition 2.4.}
{\sl
If $U:{\mathcal P}\to{\mathcal V}({\mathcal H})$ is a generalized projective representation, then
the relations (1)-(8) imply that the following equalities hold.
$$
[P_0^2-(P_1^2+P_2^2+P_3^2),U_g]=\nop\,,\eqno(2)
$$
$$
[W_0^2
-(W_1^2+W_2^2+W_3^2),U_g]=\nop\,,\eqno(3)
$$
for all $g\in{{\mathcal P}}$, including $\trev$ and $\srev$.
}
\subsection{Classifying irreducible representations of $\mathcal P$}
Spectral properties of the self-adjoint generators are now investigated, to characterize classes of representations of $\mathcal P$.
Since by (1.i), (1.vii) the generators $P_0$, $P_1$, $P_2$, $P_3$ of a generalized projective representation $U$ of $\mathcal P$
commute with each other,
according to spectral theory \cite{RS} a common spectral measure
$E:{\mathcal B}({\RR}^4)\to\Pi({\mathcal H})$
exists such that
$$
P_0=\int\lambda dE^{(0)}_\lambda\,,\quad P_j=\int\lambda dE^{(j)}_\lambda\,,\; j=1,2,3,\eqno(10)
$$
where\quad $E^{(0)}_\lambda=E((-\infty,\lambda]\times{\RR}^3)$,\quad
$E^{(1)}_\lambda=E(\RR\times(-\infty,\lambda]\times{\RR}^2)$,\quad
$E^{(2)}_\lambda=E({\RR}^2\times(-\infty,\lambda]\times{\RR})$,\quad
$E^{(3)}_\lambda=E({\RR}^3\times(-\infty,\lambda])$\quad
are the resolutions of the identity of the individual operators $P_0$, $P_1$, $P_2$, $P_3$.
\par
Once introduced the {\sl four-operator}
$\underbar P=(P_0,P_1,P_2,P_3)\equiv (P_0,{\bf P})$, the equalities (10) can be rewritten in the more compact form
$$
\underline P =\int \underline p\,dE{\underline p}\,,\hbox{ with }
dE_{\underline p}=dE^{(0)}_{p_0}dE^{(1)}_{p_1}dE^{(2)}_{p_2}dE^{(3)}_{p_3}\,.\eqno(11)
$$
The {\sl spectrum} of $\underbar P$ can be defined as the following closed subset of $\RR^4$.
$$
\sigma({\underbar P})=\{\underbar p=(p_0,{\bf p})\in\RR^4\mid
E(\Delta_{\underbar p})\neq\nop\hbox{ for every neighborough }\Delta_{\underbar p}\hbox{ of }
\underbar p\}\,.\eqno(12)
$$
By making use of (1), the following proposition can be proved.
\vskip.4pc\noindent
{\bf Proposition 2.5.}
{\sl
Let $U:{\mathcal P}\to{\mathcal U}({\mathcal H})$ be
a projective representation of ${\mathcal P}_+^\uparrow$; the relations (1) satisfied
by the self-adjoint generators of $U$ imply that
for every Lorentz transformation $g\in{\mathcal L}_+^\uparrow$ the following relation holds
$$
U_g\underbar PU_g^{-1}=\hbox{\rm\textsf g}(\underbar P),\eqno(13)
$$
where ${\rm\textsf g}:\RR^4\to\RR^4$ is the function that transforms any
$\underbar p\in\RR^4$ as a four-vector according to $g$. Moreover, the following statement is a straightforward implication of (13).
$$
U_gE(\Delta)U_g^{-1}=E({\textsf g}^{-1}(\Delta))\hbox{ holds for every }g\in{\mathcal L}_+^\uparrow.
\eqno(14)
$$
}
\subsubsection{Spectral properties in irreducible representations}
A generalized projective representation $U:{\mathcal P}\to{\mathcal V}({\mathcal H})$
can be reducible or not; in the case that
it is reducible, however, it must be the direct sum or the direct integral of irreducible ones \cite{BarBook}.
Therefore, to determine all possible generalized projective representations of ${\mathcal P}$
it is sufficient to identify the irreducible ones. For this reason, from now on we specialize to
{\sl irreducible} generalized projective representations of $\mathcal P$.
Hence, from Prop. 2.4 the following proposition follows.
\vskip.4pc\noindent
{\bf Proposition 2.6.}
{\sl
Properties (2), (3) imply that if a generalized projective representation $U$ of $\mathcal P$ is irreducible,
then two real numbers $\eta$, $\varpi$ exist such that the following equalities hold.
$$
(i)\;\;P_0^2-{\bf P}^2=\eta\Id\quad\hbox{and}\quad(ii)\;\;W^2\equiv W_0^2
-(W_1^2+W_2^2+W_3^2)=\varpi\Id\,.\eqno(4)
$$}
\vskip.5pc\noindent
Therefore every irreducible generalized projective representation of ${\mathcal P}$ is characterized by the real constant
$\eta,\varpi$. We restrict our attention to those irreducible representations of ${\mathcal P}$
for which $\eta>0$.
\vskip.5pc\noindent
Now we show that for an irreducible generalized projective representation of ${\mathcal P}$, characterized by specific
parameters $\eta>0$ and $\varpi$,
the spectrum $\sigma(\underbar P)$ of the four-operator $\underbar P=(P_0,{\bf P})$,
must be one of three definite subsets $S^+_\mu$, $S^-_\mu$, $S^+_\mu\cup S^-_\mu$ of $\RR^4$, where
$\mu$ denotes the {\sl positive} square root $\sqrt{\eta}$, and
$$
S^{+}_{\mu}=\{\underbar p\mid p_0^2-{\bf p}^2=\mu^2,\,p_0>0\}\,,\quad
S^{-}_{\mu}=\{\underbar p\mid p_0^2-{\bf p}^2=\mu^2,\,p_0<0\}\,.\eqno(15)
$$
\vskip.4pc\noindent
{\bf Proposition 2.7.}
{\sl
If $U:{\mathcal P}\to {\mathcal V}({\mathcal H})$ is an irreducible generalized projective representation,
then there are only the following mutually exclusive possibilities for the spectrum $\sigma(\underbar P)$.
\vskip.3pc\noindent
({\bf u})\quad $\sigma(\underbar P)=S^{+}_{\mu}$, ``up'' spectrum;
\vskip.3pc\noindent
({\bf d})\quad $\sigma(\underbar P)=S^{-}_{\mu}$, ``down'' spectrum;
\vskip.3pc\noindent
({\bf s})\quad $\sigma(\underbar P)=S^{+}_{\mu}\cup S^{-}_{\mu}$, ``symmetrical'' spectrum
}
\vskip.4pc\noindent
{\bf Proof.}
Since $P_0^2-{\bf P}^2-\mu^2=\nop$, if $\underbar p\in\sigma(\underbar P)$ then $p_0^2-{\bf p}^2-\mu^2=0$ must hold,
i.e.
$$\sigma(\underbar P)\subseteq\{\underline p\mid p_0^2-{\bf p}^2-\mu^2=0\}\equiv S^{+}_{\mu}\cup S^{-}_{\mu}\,.\eqno(16)$$
On the other hand,
if $\underbar p\in\sigma(\underbar P)$, then according to spectral theory $\textsf g(\underbar p)\in\sigma(\textsf g(\underbar P))$ holds
for all $g\in{\mathcal L}_+^\uparrow$,
of course; but $\textsf g(\underbar P)=U_g\underbar PU_g^{-1}$ by Prop. 2.5; therefore,
$\underbar p\in\sigma(\underbar P)$ if and only if $\textsf g(\underbar p)\in\sigma(\underbar P)$
because $\underbar P$ and $U_g\underbar PU_g^{-1}$ have the same spectra.
Hence,
$$
\sigma(\underbar P)\cap S^{+}_{\mu,\varpi}\neq\emptyset\hbox{ implies }
S^{+}_{\mu,\varpi}\subseteq\sigma(\underbar P)\,,\;\hbox{ and }\;
\sigma(\underbar P)\cap S^{-}_{\mu,\varpi}\neq\emptyset\hbox{ implies }
S^{-}_{\mu,\varpi}\subseteq\sigma(\underbar P)\,
.\eqno(17)$$
Since $\sigma(\underbar P)\neq\emptyset$, (16) and (17) imply that only one of the three cases
({\bf u}), ({\bf d}) or ({\bf s}) can occur.
\hfill{$\bullet$}
\vskip.5pc\noindent
In the case ({\bf s}) the restriction
$U:{\mathcal P}_+^\uparrow\to{\mathcal U}({\mathcal H})$ is always reducible, namely $U\mid_{{\mathcal P}_+^\uparrow}$ is reduced by
the projection operators $E^{+}=\int_{S^+_\mu}dE_{\underline p}\equiv\int_\mu^\infty p_0dE_{p_0}^{(0)}$
and $E^{-}=\int_{S^-_\mu}dE_{\underline p}\equiv\int_{-\infty}^{-\mu} p_0dE_{p_0}^{(0)}$, with
ranges ${\mathcal M}^+=E^+{\mathcal H}$ and ${\mathcal M}^-=E^-{\mathcal H}$, respectively.
We prove this statement in the following Proposition.
\vskip.5pc\noindent
{\bf Proposition 2.8.}
{\sl
In an irreducible generalized projective representation $U:{\mathcal P}\to{\mathcal V}({\mathcal H})$
the relation $[E^+,U_g]=\nop$ holds for all $g\in{\mathcal P}_+^\uparrow$ .}
\vskip.4pc\noindent
{\bf Proof.}
In the case ({\bf u}) and ({\bf d}), the statement is trivial because $E^\pm=\int_{S_\mu^\pm}dE_{\underline p}\equiv\int_{\sigma(\underline P}dE_{\underline p}=\Id$
Then we suppose that $\sigma(\underline P)=S_\mu^+\cup S_\mu^-$.
Since $E^+=\chi_{S^+_\mu}(\underline P)$,
where $\chi_{S^+_\mu}$ is the characteristic functional of $S^+_\mu$, the relations (1.i),(1.vii) imply that
$E^+$ commutes with $P_0$ and with all $P_j$. Therefore it remains to show that ${\mathcal M}^+$ is left invariant by
$U_g$, for every $g\in{\mathcal L}_+^\uparrow$.
If $\psi\in{\mathcal M}^+$, then for every $g\in{\mathcal L}_+^\uparrow$ we have
$U_g\psi=U_g\int_{S^+_\mu}dE_{\underline p}\psi=
\int_{{S^+_\mu}}dU_gE_{\underline p}{U_g}^{-1}(U_g\psi)
=\int_{S^+_\mu}dE_{{\textsf g}^{-1}({\underline p})}(U_g\psi)$, by Prop. 2.5.
The last integral is a vector of ${\mathcal M}^+=
E^+\mathcal H$, because ${\textsf g}^{-1}(\underline p)\in S_\mu^+$ if $\underline p\in S_\mu^+$ for $g\in{\mathcal L}_+^\uparrow$.\hfill{$\bullet$}
\vskip.5pc\noindent
When $\sigma(\underline P)=S^{+}_{\mu}\cup S^{-}_{\mu}$ every vector $\psi\in\mathcal H$ can be represented
as a column vector $\psi\equiv\left[\begin{matrix}\psi^+\cr \psi^-\end{matrix}\right]$, where $\psi^+=E^+\psi$ and
$\psi^-=E^-\psi$. Coherently with such a representation, any linear or anti-linear operator $A$ is represented by a matrix
$A\equiv\left[\begin{matrix}A_{11}&A_{12}\cr A_{21}&A_{22}\end{matrix}\right]$, where $A_{11}=E^+AE^+$, $A_{1}=E^+AE^-$, $A_{21}=E^-AE^+$
and $A_{22}=E^-AE^-$, in such a way that $A\psi=\left[\begin{matrix}A_{11}&A_{12}\cr A_{21}&A_{22}\end{matrix}\right]\left[\begin{matrix}\psi^+\cr \psi^-\end{matrix}\right]$. Prop.2.8 implies that, according to such a representation,
the generators $P_0$, $P_j$, $J_k$, $K_j$ have a diagonal form; by (5)-(8),
$\SREV$ and $\TREV$ have diagonal representation only if are unitary and anti-unitary, respectively.
\subsubsection{Characterization by $\TREV$ and $\SREV$ unitarity}
Now we show how each of the possibilities for $\sigma(\underline P)$ established by Prop.2.7 is characterizable
according to the unitary or anti-unitary character of the time reversal and the space inversion operators
$\TREV$ and $\SREV$.
\vskip.5pc\noindent
{\bf Lemma 2.1.}
{\sl Let $T$ be a unitary or anti-unitary operator, and let $A$ be a self-adjoint operator with spectral measure
$E^A:{\mathcal B}(\RR)\to \Pi({\mathcal H})$. If
$TAT^{-1}=f(A)$, where $f$ is a continuous bijection of $\RR$, then
$TE^A(\Delta)T^{-1}=E^A(f^{-1}(\Delta))$, for all $\Delta\in{\mathcal B}(\RR)$.}
\vskip.4pc\noindent
{\bf Proof.}
We recall that if $T$ is unitary or anti-unitary, then an operator $Q$ is a projection operator if and only if $TQT^{-1}$ is a projection operator.
This implies that if $\Delta\to E^A(\Delta)$ is the spectral measure of $A$, then $\Delta\to F(\Delta)=TE_A(\Delta)T^{-1}$ is
the spectral measure of $f(A)$. Now, let us define $\tilde\Delta=f^{-1}(\Delta)$. If $\pi(-a,a)$ is a partition of the interval
$[-a,a]$ formed by sub-intervals $\tilde\Delta_j$ with $\tilde\lambda_j\in\tilde\Delta_j$, then according to spectral theory we can write
\par\noindent
$f(A)=\lim_{{\Vert\pi(-a,a)\Vert\to 0}\atop {a\to\infty}}\sum_j f(\tilde\lambda_j)E(\tilde\Delta_j)=
\lim_{{\Vert\pi(-a,a)\Vert\to 0}\atop {a\to\infty}}\sum_j \lambda_j E(f^{-1}(\Delta_j))$\par\quad
$=\lim_{{\Vert\pi(-a,a)\Vert\to 0}\atop {a\to\infty}}\sum_j \lambda_j F(\Delta_j)$, where $\lambda_j=f(\tilde\lambda_j)\in\Delta_j$.
\par\noindent
Therefore, for the uniqueness of the spectral measure $F$ of $f(A)$ we have $F(\Delta)=E(f^{-1}(\Delta))=TE(\Delta)T^{-1}$.
\hfill{$\bullet$}
\vskip.5pc\noindent
{\bf Proposition 2.9.}
Let $U:({\mathcal P})\to{\mathcal U}({\mathcal H})$ be an irreducible generalized projective representation.
{\sl If {$\TREV$} is anti-unitary and {\rm $\SREV$} is unitary, then either
$\sigma(\underbar P)=S_\mu^+$ or $\sigma(\underbar P)=S_\mu^-$, and hence $\sigma(\underbar P)=S_\mu^+\cup S_\mu^-$
cannot occur.}
\vskip.4pc\noindent
{\bf Proof.}
We show that under the hypotheses ${\mathcal M}^+$ and ${\mathcal M}^-$ are invariant under both $\TREV$ and $\SREV$;
therefore, since ${\mathcal M}^+$ and ${\mathcal M}^-$ are invariant under the restriction $U\mid_{{\mathcal P}_+^\uparrow}$ according to Prop. 2.7,
they are invariant under the whole $U$.
If $\sigma(\underline P)=S_\mu^+\cup S_\mu^-$ held, then ${\mathcal M}^+$ would be a {\sl proper}
subspace of $\mathcal H$, so that $U$ would be reducible, in contradiction with the hypothesis of irreducibility.
\par
Now we prove the invariance of ${\mathcal M}^+$.
According to (8) the relation $\TREV P_0\TREV^{-1}=P_0$ holds when $\TREV$ is anti-unitary;
therefore Lemma 2.1 applies with $A=P_0$, $T=\TREV$ and $f$ the identity function, so that
$\TREV E^{(0)}(\Delta)\TREV^{-1}=E^{(0)}(\Delta)$ holds. This implies that
if $\psi$ is any non vanishing vector in ${\mathcal M}^+$, then
$\TREV\psi=\int_\mu^\infty d(\TREV E_{p_0}^{(0)}\TREV^{-1})\TREV\psi
=\int_\mu^\infty d\,E^{(0)}_{p_0}\TREV\psi$. Thus $\TREV\psi$ is a vector in ${\mathcal M}^+$.
\par
This argument can be repeated with $\SREV$ instead of $\TREV$, to deduce, by making use of (5), that $\SREV\psi\in{\mathcal M}^+$
for all $\psi\in{\mathcal M}^+$.
The invariance of ${\mathcal M}^{-}$ is proved quite similarly.\hfill{$\bullet$}
\vskip.5pc\noindent
{\bf Proposition 2.10.}
{\sl If $\TREV$ is unitary then $\sigma(\underline P)=
S_\mu^+\cup S_\mu^-$, independently of $\SREV$.}
\vskip.4pc\noindent
{\bf Proof.}
If $\TREV$ is unitary, then
$\sigma(\TREV P_0\TREV^{-1})=\sigma(P_0)$. But $\TREV P_0\TREV^{-1}=-P_0$ holds by (7);
therefore $-\sigma(P_0)\equiv\sigma(-P_0)=\sigma(\TREV P_0\TREV^{-1})=\sigma(P_0)$, i.e.
$$
\sigma(P_0)=-\sigma(P_0).\eqno(18)
$$
Now,
in general we have $\sigma(\underline P)=S_\mu^+$ if and only if $\sigma(P_0)=[\mu,\infty)$,
$\sigma(\underline P)=S_\mu^-$ if and only if $\sigma(P_0)=(-\infty,-\mu]$, and
$\sigma(\underline P)=S_\mu^+\cup S_\mu^-$ if and only if $\sigma(P_0)=(-\infty,-\mu]\cup[\mu,\infty)$;
Equation (18) holds only if $\sigma(P_0)=(-\infty,-\mu]\cup[\mu,\infty)$; thus
$\sigma(\underline P)=S_\mu^+\cup S_\mu^-$.\hfill{$\bullet$}
\vskip.5pc\noindent
{\bf Proposition 2.11.}
{\sl
If $\SREV$ is anti unitary then $\sigma(\underline P)=S_\mu^+\cup S_\mu^-$, independently of $\TREV$.}
\vskip.4pc\noindent
{\bf Proof.}
Since $\sigma(P_0)$ is not empty, at least one of the projection operators
$E^+=E^{(0)}([\mu,\infty))$ or $E^-=E^{(0)}((-\infty,-\mu])$
must be different from the null operator $\nop$.
\par
Let us suppose that $E^{(0)}([\mu,\infty))\neq\nop$, so that $S_\mu^+\subseteq\sigma(\underline P)$. Since
$\SREV P_0\SREV^{-1}=-P_0$ holds by (6), Lemma 2.1 can be applied to deduce
$\SREV E^{(0)}([\mu,\infty))\SREV^{-1}=E^{(0)}((-\infty,-\mu])$; hence $E^{(0)}((-\infty,-\mu])$ is a non null projection operator
because $E^{(0)}([\mu,\infty))$ is non-null and $\SREV$ is anti-unitary. This means that
$\sigma(P_0)\cap(-\infty,-\mu]$ is not empty, that is to say
that $\sigma(\underline P)\cap S_\mu^-\neq\emptyset$; therefore, according to Prop. 2.7,
$\sigma(\underline P)=S_\mu\cup S_\mu^-$.
\par
In the case that $E^{(0)}((-\infty,-\mu])\neq\nop$ the argument is easily adapted to
reach the same conclusion.\hfill{$\bullet$}
\section{Irreducible $U$ with $U^+\mid_{{\mathcal P}_+^\uparrow}$ irreducible}
The present section and the next one are devoted to identify the possible structures of the irreducible generalized projective representations
of the Poincar\'e group.
Let us highlight some results of the previous section.
Given an irreducible generalized projective representation $U:{\mathcal P}\to{\mathcal U}({\mathcal H})$, only one of three mutually exclusive cases can occur:
$$
({\bf u})\;\;\sigma(\underline P)=S_\mu^+,\quad({\bf d})\;\;\sigma(\underline P)=S_\mu^-,\quad({\bf s})\;\;\sigma(\underline P)=S_\mu^+\cup S_\mu^-.
$$
The cases $\sigma(\underline P)=S_\mu^+$ or $\sigma(\underline P)=S_\mu^-$ occur if and only if $\TREV$ is anti-unitary and $\SREV$ is unitary.
\par\noindent
Any other combination, that is to say
$\TREV$ unitary and $\SREV$ unitary, $\TREV$ unitary and $\SREV$ anti-unitary or
$\TREV$ anti-unitary and $\SREV$ anti-unitary,
characterizes irreducible generalized projective representations of $\mathcal P$ with symmetrical spectrum $\sigma(\underline P)=S_\mu^+\cup S_\mu^-$.
\par\noindent
Moreover, in the case of symmetrical spectrum $\sigma(\underline P)=S_\mu^+\cup S_\mu^-$,
the restriction $U\mid_{{\mathcal P}_+^\uparrow}$ is reduced by $E^+$ into
$U^+\mid_{{\mathcal P}_+^\uparrow}=E^+U\mid_{{\mathcal P}_+^\uparrow}E^+$
\;and \;$U^-\mid_{{\mathcal P}_+^\uparrow}=E^-U\mid_{{\mathcal P}_+^\uparrow}E^-$.
\par\noindent
If $\sigma(\underline P)=S_\mu^+$ (resp., $\sigma(\underline P)=S_\mu^-$),
then $U\mid_{{\mathcal P}_+^\uparrow}=U^+\mid_{{\mathcal P}_+^\uparrow}$ (resp., $U\mid_{{\mathcal P}_+^\uparrow}=U^-\mid_{{\mathcal P}_+^\uparrow}$).
\vskip.5pc
An effective help in accomplishing our task will provided just by the
investigation of the reductions $U^+\mid_{{\mathcal P}_+^\uparrow}$ or $U^-\mid_{{\mathcal P}_+^\uparrow}$.
In general, even if the ``mother'' irreducible generalized projective representation $U$ is irreducible, the reductions
$U^+\mid_{{\mathcal P}_+^\uparrow}$ or $U^-\mid_{{\mathcal P}_+^\uparrow}$
can be reducible or not.
In the present section we completely identify the possible irreducible generalized projective representations $U$ of $\mathcal P$ for which
$U^+\mid_{{\mathcal P}_+^\uparrow}$ and $U^-\mid_{{\mathcal P}_+^\uparrow}$ are {\sl irreducible}.
In doing so we shall identify, besides the well known irreducible generalized projective representations $U$ of $\mathcal P$
with $\TREV$ anti-unitary and $\SREV$ unitary, or with $\TREV$ unitary and $\SREV$ unitary, also those with $\TREV$ anti-unitary and $\SREV$ anti-unitary,
or with $\TREV$ unitary and $\SREV$ anti-unitary, which are not taken into account in the literature about elementary particles theory.
\subsection{The case $\sigma(\underline P)=S_\mu^\pm$ with $U^\pm\mid_{{\mathcal P}_+^\uparrow}$ irreducible}
The irreducible generalized  projective representation of ${\mathcal P}$ with $\sigma(\underline P)=S_\mu^+$ and
whose restriction to ${\mathcal P}_+^\uparrow$ is irreducible are well known \cite{W2},\cite{c13}.
For each allowed pair $\varpi$, $\eta=\mu^2>0$ of the parameters characterizing
the irreducible generalized projective representation of $\mathcal P$, {\sl modulo unitary isomorphisms} there is only one irreducible projective representation
of ${\mathcal P}_+^\uparrow$ with $\sigma(\underline P)=S_\mu^+$ and only one with $\sigma(\underline P)=S_\mu^-$, that we briefly present.
The allowed value of $\varpi$ are $\varpi=\mu s(s+1)$, where $s\in\frac{1}{2}\NN$.
Following a common practice, $s$ will be called {\sl spin parameter} or simply {\sl spin}.
Once fixed $s$ and $\mu$, the Hilbert space of the projective representation is the space $L_2(\RR^3,\CC^{2s+1},d\nu)$ of functions
$\psi:\RR^3\to \CC^{2s+1}$,
${\bf p}\to\psi({\bf p})$, square integrable with respect to the measure
$d\nu({\bf p}=\frac{dp_1dp_2dp_3}{\sqrt{\mu^2+{\bf p}^2}}$.
\subsubsection{The case $\sigma(\underline P)=S_\mu^+$}
Fixed $\mu$ and $s$, for the irreducible projective representation with $\sigma(\underline P)=S_\mu^+$ the following statements hold.
\begin{itemize}
\item[--]
The generators $P_j$ are the multiplication operators defined by
$(P_j\psi)({\bf p})=p_j\psi({\bf p})$, and as consequence
\item[--]
$(P_0\psi)({\bf p})=p_0\psi({\bf p})$ where $p_0=+\sqrt{\mu^2+{\bf p}^2}$,
because $P_0$ has a positive spectrum;
\item[--]
the generators $J_k$ are given by $J_k=i\left(p_l\frac{\partial}{\partial p_j}-p_j\frac{\partial}{\partial p_l}\right)+S_k$,
$(k,l,j)$ being a cyclic permutation of $(1,2,3)$,
where $S_1,S_2,S_3$ are the self-adjoint generators of an irreducible projective representation $L:SO(3)\to\CC^{2s+1}$
such that $S_1^2+S_2^2+S_3^2=s(s+1)\Id$;
hence, they can be fixed to be the three spin operators of $\CC^{2s+1}$;
\item[--]
the generators $K_j$ are given by
$K_j=ip_0\frac{\partial}{\partial p_j}-\frac{({\bf S}\land {\bf p})_j}{\mu+p_0}$;
\item[--]
the unitary space inversion operator and the anti-unitary time reversal operator are
$$
\SREV=\Upsilon,\quad\hbox{and}\quad\TREV=\tau{\mathcal K}\Upsilon
\eqno(19)
$$
\item[]
where
$\Upsilon$ is the unitary operator defined by
$(\Upsilon\psi)({\bf p})=\psi(-{\bf p})$, and
\item[]
where
$\tau$ is a unitary matrix of \,$\CC^{2s+1}$ such that $\tau {\overline S_j}\tau^{-1}= -{S_j}$, for all $j$;
such a matrix always exists and it is unique up a complex factor of modulus 1;
moreover,
if $s\in\NN$ then $\tau$ is symmetric and $\tau\overline\tau=1$, while if $s\in\left(\NN+\frac{1}{2}\right)$ then $\tau$ is
anti-symmetric and $\tau\overline\tau=-1$ \cite{c13};
\item[-]
$\mathcal K$ is the anti-unitary complex conjugation operator defined by ${\mathcal K}\psi({\bf p})=\overline{\psi({\bf p})}$.
\end{itemize}
\subsubsection{The case $\sigma(\underline P)=S_\mu^-$}
For the irreducible projective representation with $\sigma(\underline P)=S_\mu^-$, the following {\sl symmetrical}
statements hold.
\begin{itemize}
\item[--]
the generators $P_j$ are the multiplication operators defined by
$(P_j\psi)({\bf p})=p_j\psi({\bf p})$, and as consequence
\item[--]
$(P_0\psi)({\bf p})=-p_0\psi({\bf p})$,
because $P_0$ has a negative spectrum;
\item[--]
the generators $J_k$ are give
n by $J_k=i\left(p_l\frac{\partial}{\partial p_j}-p_j\frac{\partial}{\partial p_l}\right)+S_k$,
$(k,l,j)$ being a cyclic permutation of $(1,2,3)$;
\item[--]
the generators $K_j$ are given by
$K_j=-ip_0\frac{\partial}{\partial p_j}+\frac{({\bf S}\land {\bf P})_j}{\mu+p_0}$;
\item[--]
the space inversion -- unitary -- and time reversal -- anti-unitary --  operators are
$\SREV=\Upsilon$ and $\TREV=\tau{\mathcal K}\Upsilon$.
\end{itemize}
\subsection{The case $\sigma(\underline P)=S_\mu^+\cup S_\mu^-$ with $U^+\mid_{{\mathcal P}_+^\uparrow}$ irreducible}
Now we establish results that allow us to identify the irreducible generalized projective representations
with $U^+\mid_{{\mathcal P}_+^\uparrow}$ irreducible.
Prop. 2.10 and Prop. 2.11 imply that the case $\sigma (\underline P)=S_\mu^+\cup S_\mu^-$ can occur only if
$\TREV$ is unitary or $\SREV$ is anti-unitary. Moreover, according to Prop. 2.8, if $U:{\mathcal P}\to{\mathcal V}({\mathcal H})$ is irreducible
with $\sigma(\underline P)=S_\mu^+\cup S_\mu^-$, then $U\mid_{{\mathcal P}_+^\uparrow}$ is always reduced by
$E^+=E(S_\mu^+)\equiv E^{(0)}[\mu,\infty)=\chi_{[\mu,\infty)}(P_0)$ and
$E^-=E(S_\mu^-)\equiv E^{(0)}(-\infty,-\mu]=\chi_{(-\infty,-\mu]}(P_0)$, so that
we can write $U_g=E^+U_gE^++E^-U_gE^-$, for all $g\in{\mathcal P}_+^\uparrow$.
\par
Each of the two components $U^+:{\mathcal P}_+^\uparrow\to{\mathcal U}({\mathcal H})$, $g\to U_g^+=E^+U_gE^+$ and
$U^-:{\mathcal P}_+^\uparrow\to{\mathcal U}({\mathcal H})$, $g\to U_g^-=E^-U_gE^-$
can be reducible or not, in its turn.
The following proposition entails that the reducibility of $U^+\mid_{{\mathcal P}_+^\uparrow}$ is equivalent to the reducibility of
$U^-\mid_{{\mathcal P}_+^\uparrow}$.
\vskip.5pc\noindent
{\bf Proposition 3.1.}
{\sl
Let $U:{\mathcal P}\to{\mathcal V}({\mathcal H})$ irreducible with $\sigma(\underline P)=S_\mu^+\cup S_\mu^-$.
If $F_+$ is a projection operator that reduces $U^+\mid_{{\mathcal P}_+^\uparrow}$, then the following statements hold.
\begin{itemize}
\item[(i)]
In the case that $\TREV$ is unitary, the projection operator $F_-^\trev=\TREV F_+\TREV$ reduces
$U^-({\mathcal P}_+^\uparrow)$ and $F^\trev=F_++ F_-^\trev$ reduces $U\mid_{{\mathcal P}_+}$;
\item[(ii)]
in the case that $\SREV$ is anti-unitary, the projection operator $F_-^\srev=\SREV F_+\SREV$ reduces
$U^-({\mathcal P}_+^\uparrow)$ and $F^\srev=F_++ F_-^\srev$ reduces $U\mid_{{\mathcal P}^\uparrow}$.
\end{itemize}}
\vskip.4pc\noindent
{\bf Proof.}
If $\TREV$ is unitary, then
$\TREV^{-1}=\TREV$ and $\TREV P_0\TREV=-P_0$ follow from (7); this implies
$\TREV E^+\TREV=\TREV\chi_{[\mu,\infty)}(P_0)\TREV=\chi_{(-\infty,-\mu]}=E^-$ by Lemma 2.1.
If $F_+$ is a projection operator that reduces $U^+\mid_{{\mathcal P}_+^\uparrow}$, and hence $\nop<F_+< E^+$,
then $F_-^\trev E^-=(\TREV F_+\TREV)E^-=(\TREV F_+\TREV)\TREV E^+\TREV=\TREV F_+ E^+\TREV=\TREV F_+ \TREV$ since $F_+< E^+$.
Therefore, $\nop F_-^\trev< E_-$ is satisfied.
Now we show that $[F_-^\trev,P_0^-]=[F_-^\trev,P_j^-]=[F_-^\trev,K_j^-]=[F_-^\trev,J_k^-]=\nop$, i.e. that
$F_-^\trev$ reduces $U^-\mid_{ {\mathcal P}_+^\uparrow}$.
Since $P_0^-=E^-P_0E^-$,
we have\par\noindent
$P_0^-F_-^\trev=P_0^-F_-^\trev E^-=
E^-P_0E^-\TREV F_+\TREV E^-=$
$E^-P_0\TREV E^+\TREV \TREV F_+\TREV E^-=$
$-E^-\TREV P_0 E^+ F_+ \TREV E^-=$
$-E^-\TREV E^+ P_0 F_+ \TREV E^-=$
$-E^-\TREV E^+F_+ P_0\TREV E^-=$
$-E^-\TREV  F_+ P_0\TREV E^-=$
$E^-\TREV F_+\TREV  P_0 E^-=$
$E^-F_-^\trev P_0 E^-=$
$F_-^\trev E^- P_0E^-=$
$F_-^\trev P_0^-$.
\par\noindent
A similar derivation shows that
$[F_-^\trev,P_j]=[F_-^\trev,K_j^-]=[F_-^\trev,J_k^-]=\nop$; therefore $F_-^\trev$ reduces $U^-\mid_{{\mathcal P}_+^\uparrow}$.
Now we see that $F^\trev=F_++F_-^\trev$ reduces $U\mid_{{\mathcal P}_+}$. The equalities
$F^\trev P_0=(F_++F_-^\trev)P_0=P_0(F_++F_-^\trev)=P_0F^\trev$ immediately follow from
$P_0=E^+P_0 E^++E^-P_0E^-$ and $F_-^\trev E^-=F_-^\trev$, $F_+E^+=F_+$, $F_+E^-=F_-^\trev E^+=\nop$;
similarly,
$[F_-^\trev,P_j]=[F_-^\trev,J_k]=[F_-^\trev,K_j]=\nop$ hold. Hence, $F^\trev$ reduces $U\mid_{{\mathcal P}_+^\uparrow}$.
Moreover, $F^\trev\TREV=F_+\TREV+F_-^\trev\TREV=\TREV\TREV F_+\TREV+\TREV F_+\TREV\TREV=\TREV F_-^\trev+\TREV F_+=\TREV F^\trev$.
Therefore, $F^\trev$ reduces also $U\mid_{{\mathcal P}_+}$.
A quite similar argument proves statement (ii).
\hfill{$\bullet$}
\vskip.5pc\noindent
{\bf Corollary.}
{\sl Under the hypotheses of Prop. 3.1, $U^+({\mathcal P}_+^\uparrow)$ is reducible if and only if  $U^-({\mathcal P}_+^\uparrow)$ is reducible.}
\vskip.5pc\noindent
Prop. 3.1 and its corollary indicate that the irreducible generalized projective representations of $\mathcal P$
can be classified according to the reducibility of $U^+\mid_{{\mathcal P}_+^\uparrow}$.
\subsubsection{Hilbert space and self-adjoint generators}
In the case that $U^+\mid_{{\mathcal P}_+^\uparrow}$ is irreducible, with $\sigma(\underline P)=S_\mu^+\cup S_\mu^-$,
according to Prop. 2.8 the restriction $U:{\mathcal P}_+^\uparrow\to{\mathcal U}({\mathcal H})$ must be the direct sum of
$U^+:{\mathcal P}_+^\uparrow\to{\mathcal U}({\mathcal H}^+)$ and $U^-:{\mathcal P}_+^\uparrow\to{\mathcal U}({\mathcal H}^-)$,
where ${\mathcal H}^+=E^+({\mathcal H})$, ${\mathcal H}^-=E^-({\mathcal H})$ and ${\mathcal H}^+\oplus{\mathcal H}^-={\mathcal H}$;
according to Prop. 3.1, both $U^+\mid_{{\mathcal P}_+^\uparrow}$ and $U^-\mid_{{\mathcal P}_+^\uparrow}$
are irreducible projective representations of ${\mathcal P}_+^\uparrow$. Since $\sigma(P_0^+)=[\mu,\infty)$
(resp., $\sigma(P_0^-)=(-\infty,-\mu]$), the reduced projective representation $U^+\mid_{{\mathcal P}_+^\uparrow}$ (resp., $U^-\mid_{{\mathcal P}_+^\uparrow}$) is
unitarily isomorphic to the projective representation $U:{\mathcal P}_+^\uparrow\to {\mathcal U}(L_2(\RR^3,\CC^{2s+1},d\nu))$ with
$\sigma(\underline P)=S_\mu^+$
(resp., with $\sigma(\underline P)=S_\mu^-$) described in sect. 3.1, with the same characterizing parameters $\mu$ and $s$ of the
unrestricted irreducible generalized projective representation $U$.
Accordingly, there are two unitary mappings $W^+:{\mathcal H}^+\to L_2(\RR^3,\CC^{2s+1},d\nu)$, and $W^-:{\mathcal H}^-\to L_2(\RR^3,\CC^{2s+1},d\nu)$
such that the reduced generators in $W^+({\mathcal H}^+)\equiv L_2(\RR^3,\CC^{2s+1},d\nu)$ are
$P_0^+=W^+(E^+P_0E^+)$, that acts as a multiplication by $p_0$, i.e.
$(P_0^+\phi)({\bf p})=p_0({\bf p})\phi({\bf p})=\sqrt{\mu^2+{\bf p}^2}\phi({\bf p})$; $P_j^+=p_j$;
$J_k^+={\textsf j}_k=i\left(p_l\frac{\partial}{\partial p_j}-p_j\frac{\partial}{\partial p_l}\right)+S_k$;
$K_j^+={\textsf k}_j=ip_0\frac{\partial}{\partial p_j}-\frac{({\bf S}\land {\bf P})_j}{\mu+p_0}$.
Symmetrically,
the reduced generators in $W^-({\mathcal H}^-)\equiv L_2(\RR^3,\CC^{2s+1},d\nu)$ are
$P_0^-=W^-(E^-P_0E^-)=-p_0$; $P_j^-=p_j$;
$J_k^-={\textsf j}_k=i\left(p_l\frac{\partial}{\partial p_j}-p_j\frac{\partial}{\partial p_l}\right)+S_k$;
$K_j^-=-{\textsf k}_j=-ip_0\frac{\partial}{\partial p_j}+\frac{({\bf S}\land {\bf p})_j}{\mu+p_0}$.
\par
Hence we have proved that, modulo unitary isomorphisms, the Hilbert space the representation
is $L_2(\RR^3,\CC^{2s+1},d\nu)\oplus L_2(\RR^3,\CC^{2s+1},d\nu)$.
It is convenient to represent each vector $\psi\in{\mathcal H}=E^+\psi+E^-\psi$ as a column vector
$\psi=\left[\begin{matrix}\psi^+\cr \psi^-\end{matrix}\right]$, where $\psi^+=W^+(E^+\psi)$ and $\psi^-=W^+(E^-\psi)$;
in such a representation the generators of $U\mid_{{\mathcal P}_+^\uparrow}$ are
the following operators, known as the {\sl canonical form}.
$$
P_0=\left[\begin{matrix}p_0&0\cr 0&-p_0\end{matrix}\right],\quad
J_k=\left[\begin{matrix}{\textsf j}_k&0\cr 0&{\textsf j}_k\end{matrix}\right],\quad
K_j=\left[\begin{matrix}{\textsf k}_j&0\cr 0&-{\textsf k}_j\end{matrix}\right].\eqno(20)
$$
\subsubsection{Time reversal and space inversion operators}
Since
$\sigma(\underline P)=S_\mu^+\cup S_\mu^-$,
the time reversal operators $\TREV$ must be unitary or the space inversion operator $\SREV$
must be anti-unitary.
In the case in which both $\TREV$ {\sl and} $\SREV$ are
unitary their explicit form is well known, up a complex factor of modulus 1 \cite{c13}.
$$
\TREV=\left[\begin{matrix}0&1\cr 1&0\end{matrix}\right]\,;\qquad \SREV=\Upsilon\left[\begin{matrix}1&0\cr 0&1\end{matrix}\right]
\quad\hbox{or}\quad\SREV=\Upsilon\left[\begin{matrix}1&0\cr 0&-1\end{matrix}\right]\,.
\eqno(21)
$$
(In the matrices (21) ``$1$'' and ``$0$'' denote the identity and null operators of\;  $\CC^{2s+1}$. This notation is adopted throughout the paper,
whenever it does not cause confusions)
\vskip.5pc\noindent
In fact, irreducible generalized projective representations with $\TREV$ anti-unitary or $\SREV$ anti-unitary do exist,
as we show after the following Prop.3.2.
\vskip.5pc\noindent
{\bf Proposition 3.2.}
{\sl
Let $U:{\mathcal P}\to{\mathcal V}({\mathcal H})$ be an irreducible generalized projective representation of $\mathcal P$,
with $U^+({\mathcal P}_+^\uparrow)$ irreducible. The following statements hold.
\begin{itemize}
\item[i)]
If $\TREV$ is anti-unitary then
\item[]
$\TREV=\tau{\mathcal K}\Upsilon\left[\begin{matrix}1&0\cr 0&e^{i\theta}\end{matrix}\right]$;
hence, $\TREV$ can be taken to be $\TREV=\tau{\mathcal K}\Upsilon\left[\begin{matrix}1&0\cr 0&1\end{matrix}\right]$
up a complex factor of modulus 1;
\item[] in particular,
$s\in\NN$ implies $\TREV^2=\Id$ and $s\in(\NN+\frac{1}{2})$ implies $\TREV^2=-\Id$;
\item[ii)]
if $\SREV$ is anti-unitary then \item[]
$\SREV=\left[\begin{matrix}0&\tau\cr \tau&0\end{matrix}\right]{\mathcal K}$ when
$\SREV^2=\Id$ and $s\in\NN$, or when $\SREV^2=-\Id$ and $s\in(\NN+\frac{1}{2})$,
\item[]
$\SREV=\left[\begin{matrix}0&\tau\cr -\tau&0\end{matrix}\right]{\mathcal K}$ when
$\SREV^2=-\Id$ and $s\in\NN$, or when $\SREV^2=\Id$ and $s\in(\NN+\frac{1}{2})$.
\end{itemize}}
\noindent
{\bf Proof.}
Since $\TREV$ is anti-unitary,
the operator $\hat T=\tau{\mathcal K}\Upsilon\TREV\equiv\left[\begin{matrix} T_{11}&T_{12}\cr T_{21}&T_{22}\end{matrix}\right]$
is unitary, and $\TREV={\mathcal K}\Upsilon\tau\hat T$. Now, (8), $\tau^{-1}\overline{S_k}\tau=-S_k$ and the properties
\par\noindent
$\Upsilon p_j=-p_j\Upsilon$, $\Upsilon\frac{\partial}{\partial p_j}=-\frac{\partial}{\partial p_j}\Upsilon$, $\Upsilon^2=\Id$,
${\mathcal K}p_j=p_j{\mathcal K}$, ${\mathcal K}\frac{\partial}{\partial p_j}=\frac{\partial}{\partial p_j}{\mathcal K}$,
${\mathcal K}\Upsilon=\Upsilon{\mathcal K}$, ${\mathcal K}^2=\Id$
\par\noindent
imply $[\hat T,P_j]=\nop$, $[\hat T,P_0]=\nop$, $[\hat T,J_k]=\nop$. The first two equalities imply that
$\hat T=\left[\begin{matrix} T_1({\bf p})&0\cr 0& T_2({\bf p})\end{matrix}\right]$, where
$T_m({\bf p})$ is a $(2s+1)\times(2s+1)$ matrix for every ${\bf p}\in\RR^3$, so that $[T_m({\bf p}),p_j]=\nop$; the third equality implies
$[T_m({\bf p}),{\textsf j}_k]=\nop$. Then, since $p_1,p_2,p_3,{\textsf j}_1,{\textsf j}_2,{\textsf j}_3$
are the generators of an irreducible projective representation of $\mathcal E$ in the Hilbert space $L_2(\RR^3,\CC^{2s+1},d\nu)$,
the relations
$[T_m({\bf p}),p_j]=\nop$ and
$[T_m({\bf p}),{\textsf j}_k]=\nop$ imply
$T_m({\bf p})=t_m\Id$, so that
$\hat T=\left[\begin{matrix} t_1&0\cr 0& t_2\end{matrix}\right]$, with $t_1,t_2$ constant. Hence we can set
$\TREV={\mathcal K}\Upsilon\tau^{-1}\left[\begin{matrix} 1&0\cr 0& e^{i\theta}\end{matrix}\right]$. Since this last is
$\pm\tau{\mathcal K}\Upsilon\left[\begin{matrix} 1&0\cr 0& e^{i\theta}\end{matrix}\right]$, the fre phase can be chosen so that
$\TREV=\tau{\mathcal K}\Upsilon\left[\begin{matrix} 1&0\cr 0& e^{i\theta}\end{matrix}\right]$.
By transforming each operator $B$ into $WBW^{-1}$, where $W=\left[\begin{matrix} 1&0\cr 0& e^{i\frac{\theta}{2}}\end{matrix}\right]$,
$\TREV$ turns out to be transformed into $\TREV=\tau{\mathcal K}\Upsilon\left[\begin{matrix} 1&0\cr 0& 1\end{matrix}\right]$, while all generators
$P_j$, $P_0$, $J_k$, $K_j$ remain unchanged.
Accordingly, $\TREV^2=\left[\begin{matrix} \tau\overline\tau&0\cr 0& \tau\overline\tau\end{matrix}\right]$.
If $s\in\NN$, then $\tau\overline\tau=1$, so that $\TREV^2=\Id$; if $s\in(\NN+\frac{1}{2})$, then $\tau\overline\tau=-1$, so that
$\TREV^2=-\Id$. This proves (3.2.i).
\par
The proof of (3.2.ii) is carried out along quite similar lines.
\hfill{$\bullet$}
\vskip.5pc\noindent
The combination $\TREV$ anti-unitary
and $\SREV$ unitary is excluded by the condition $\sigma(\underline P)=S_\mu^+\cup S_\mu^-$.
\vskip.5pc\noindent
The combination $\TREV$ unitary and $\SREV$ unitary is already settled according to (21).
Then we check the consistency of the remaining combinations;
it is sufficient to verify that (9) is satisfied, since all other
conditions for the generators $P_J$, $P_0$, $J_k$, $K_j$ and for $\TREV$ and $\SREV$ have been already verified.
\begin{itemize}
\item[a)]
If {\sl $\TREV$ is anti-unitary and $\SREV$ is anti-unitary}, then they have the form shown by Prop. 3.2.i and
3.2.ii.
By a straightforward calculation it is verified that
condition (9) is always satisfied.
\item[b)]
If {\sl $\TREV$ is unitary and $\SREV$ is anti-unitary}, then they have the form given by (21) and Prop. 3.2.ii.
We see that (9) is always satisfied.
\end{itemize}
Thus, besides the usually considered irreducible generalized projective representations with the combination $\TREV$ unitary, $\SREV$ unitary,
also the combinations $\TREV$ unitary, $\SREV$  anti-unitary and $\TREV$ anti-unitary, $\SREV$  anti-unitary can occur.
They are completely determined by Prop. 3.2.
\section{Irreducible $U:{\mathcal P}\to{\mathcal V}({\mathcal H})$ with $U^+\mid_{{\mathcal P}_+^\uparrow}$ reducible}
In the literature about single relativistic particles only irreducible generalized projective representations
$U:{\mathcal P}\to{\mathcal V}({\mathcal H})$ with $U^\pm\mid_{{\mathcal P}_+^\uparrow}$ irreducible are considered.
This would be a correct practice
if the irreducibility of the whole $U$ implied the irreducibility of the reductions
$U^\pm\mid_{{\mathcal P}_+^\uparrow}=E^\pm U\mid_{{\mathcal P}_+^\uparrow}E^\pm$.
This is not the case.
In this section, in fact, we show that irreducible generalized projective representations $U$ of $\mathcal P$ exist such that
$U^+\mid_{{\mathcal P}_+^\uparrow}$ is reducible in the case $\sigma(\underline P)=S_\mu^\pm$, as well as in the case $\sigma(\underline P)=S_\mu^+\cup S_\mu^-$.
\subsection{The case $\sigma(\underline P)=S_\mu^+$, $\sigma(\underline P)=S_\mu^-$}
Prop. 2.8 implies that if the restriction
$U\mid_{{\mathcal P}_+^\uparrow}$ of an irreducible generalized projective representation of $\mathcal P$ is irreducible too, then either
$\sigma(\underline P)=S_\mu^+$ or $\sigma(\underline P)=S_\mu^-$.
The converse is not true; in other words, the condition $\sigma(\underline P)=S_\mu^+$
implies $U\mid_{{\mathcal P}_+^\uparrow}=U^+\mid_{{\mathcal P}_+^\uparrow}$, but does not imply
that $U^+\mid_{{\mathcal P}_+^\uparrow}\equiv U\mid_{{\mathcal P}_+^\uparrow}$ is irreducible.
In fact, now we identify irreducible generalized projective representations $U:{\mathcal P}\to{\mathcal V}({\mathcal H})$
for which $\sigma({\underline P})=S_\mu^+$ is reducible.
\par
We deal with the case $\sigma(\underline P)=S_\mu^+$;
the alternative  $\sigma(\underline P)=S_\mu^-$ can be addressed along identical lines.
We show that for any $\mu>0$ and any $s\in\frac{1}{2}\NN$ there are irreducible generalized projective
representations such that $U^+\mid_{{\mathcal P}_+^\uparrow}$ is the direct sum $U^{(1)}\oplus U^{(1)}$ of two identical projective representations
$U^{(1)}:{\mathcal P}_+^\uparrow\to {\mathcal U}({\mathcal H}^{(1)})$ and
$U^{(2)}:{\mathcal P}_+^\uparrow\to {\mathcal U}({\mathcal H}^{(2)})$.
\par
Let us consider the two irreducible projective representations
$U^{(1)}:{\mathcal P}_+^\uparrow\to {\mathcal U}({\mathcal H}^{(1)})$ and
$U^{(2)}:{\mathcal P}_+^\uparrow\to {\mathcal U}({\mathcal H}^{(2)})$ of ${\mathcal P}_+^\uparrow$ of the form
described in sect. 3.1.1, with the same pair $\mu$, $s$ of parameters that determine the representations up unitary isomorphisms.
The Hilbert space of the direct sum $U^{(1)}\oplus U^{(1)}$ is
${\mathcal H}= L_2(\RR^3,\CC^{2s+1},d\nu) \oplus  L_2(\RR^3,\CC^{2s+1},d\nu)$.
\par
It is convenient to represent
every vector $\psi=\psi_1+\psi_2$ in $\mathcal H$, with $\psi_1\in{\mathcal H}^{(1)}$ and $\psi_2\in{\mathcal H}^{(2)}$,
as the column vector $\psi\equiv \left[\begin{matrix}\psi_1\cr \psi_2\end{matrix}\right]$, so that every linear (resp., anti-linear) operator $A$ of $\mathcal H$
can be represented by a matrix $\left[\begin{matrix}A_{11}& A_{12} \cr A_{21} & A_{22}\end{matrix}\right]$, where $A_{mn}$ is a linear (resp., anti-linear)
operator from ${\mathcal H}^{(m)}\equiv L_2(\RR^3,\CC^{2s+1},d\nu)$ to ${\mathcal H}^{(n)}\equiv L_2(\RR^3,\CC^{2s+1},d\nu)$, and
$A\psi=\left[\begin{matrix}A_{11} & A_{12} \cr A_{21} & A_{22}\end{matrix}\right]  \left[\begin{matrix}\psi_1\cr \psi_2\end{matrix}\right]=
\left[\begin{matrix}
A_{11}\psi_1+A_{12}\psi_2 \cr A_{21}\psi_1+A_{22}\psi_2\end{matrix}\right]$.
Accordingly,
$P_0=\left[\begin{matrix}p_0& 0 \cr 0 & p_0\end{matrix}\right]$,
$P_j=\left[\begin{matrix}p_j& 0 \cr 0 & p_j\end{matrix}\right]$,
$J_k=\left[\begin{matrix}{\textsf j}_k& 0 \cr 0 & {\textsf j}_k\end{matrix}\right]$,
$J_j=\left[\begin{matrix}{\textsf k}_j& 0 \cr 0 & {\textsf k}_j\end{matrix}\right]$, where
${\textsf j}_k=i\left(p_l\frac{\partial}{\partial p_j}-p_j\frac{\partial}{\partial p_l}\right)+S_k$ and
${\textsf k}_j=ip_0\frac{\partial}{\partial p_j}-\frac{({\bf S}\land {\bf p})_j}{\mu+p_0}$.
\par
In such a way,
the reducible projective representation $U:{\mathcal P}_+^\uparrow\to L_2(\RR^3,\CC^{2s+1},d\nu)\oplus L_2(\RR^3,\CC^{2s+1},d\nu)$ is completely determined.
The possible extensions to the whole $\mathcal P$ are obtained by identifying the time reversal operator $\TREV$ and the space inversion operator $\SREV$.
In sections 4.1.2 and 4.1.2 we show that, while fixed $\mu$ and $s$ there is a unique possibility for $\SREV$ up to unitary equivalence,
there are inequivalent possibilities for $\TREV$. In section 4.1.3 we prove that some of these possibilities correspond to irreducible
generalized projective representations of of $\mathcal P$.
\subsubsection{Space inversion operator $\SREV$}
The condition $\sigma(\underline P)=S_\mu^+$ implies that $\TREV=\left[\begin{matrix}\TREV_{11}& \TREV_{12} \cr \TREV_{21} & \TREV_{22}\end{matrix}\right]$ is anti-unitary
and $\SREV = \left[\begin{matrix}\SREV_{11}&\SREV_{12} \cr\SREV_{21} &\SREV_{22}\end{matrix}\right]$ is unitary, according to Prop. 2.10 and 2.11.
We begin by determining $\SREV$. Relations (5) imply
$$
\SREV_{mn}p_0=p_0\SREV_{mn},\; \SREV_{mn}p_j=-p_j\SREV_{mn},\; \SREV_{mn}{\textsf j}_k={\textsf j}_k\SREV_{mn},\; \SREV_{mn}{\textsf k}_j=-{\textsf k}_j\SREV_{mn}.\eqno(22)
$$
The unitary operator
$\Upsilon$ satisfies the following relations.
$$
\Upsilon p_j=-p_j\Upsilon,\quad\Upsilon\frac{\partial}{\partial p_j}=-\frac{\partial}{\partial p_j}\Upsilon,\quad[\Upsilon, S_j]=\nop,\quad \Upsilon^2=\Id\,.
\eqno(23)
$$
Once introduced the unitary operator $\hat S=\Upsilon\SREV$, from (5), (23) and (22) we derive
\vskip.8pc\noindent
${\hat S}_{mn}p_0\equiv({\hat S}P_0)_{mn}=\Upsilon\SREV_{mn}p_0=p_0\Upsilon\SREV_{mn}=p_0\hat S_{mn}$,\hfill{(24.i)}
\vskip.7pc\noindent
${\hat S}_{mn}p_j\equiv({\hat S}P_j)_{mn}=\Upsilon\SREV_{mn}p_j=-\Upsilon p_j\SREV_{mn}=p_j\Upsilon\SREV_{mn}=p_j\hat S_{mn}$,\hfill{(24.ii)}
\vskip.7pc\noindent
${\hat S}_{mn}{\textsf j}_k\equiv({\hat S}J_k)_{mn}=\Upsilon\SREV_{mn}{\textsf j}_k=\Upsilon {\textsf j}_k\SREV_{mn}=
\Upsilon\left(ip_l\frac{\partial}{\partial p_j} -ip_j\frac{\partial}{\partial p_l}+S_k\right)\SREV_{mn}=$,\hfill{(24.iii)}
\vskip.5pc
$\qquad\quad=\left(ip_l\frac{\partial}{\partial p_j}-ip_j\frac{\partial}{\partial p_l}+S_k\right)\Upsilon\SREV_{mn}={\textsf j}_k{\hat S}_{mn}$,
\vskip.7pc\noindent
${\hat S}_{mn}k_j\equiv({\hat S}K_j)_{mn}=\Upsilon\SREV_{mn}k_j=-\Upsilon k_j\SREV_{mn}=
-\Upsilon\left(ip_0\frac{\partial}{\partial p_j}-\frac{[{\bf S}\land{\bf p}]_j}{\mu+p_0}\right)\SREV_{mn}=$,\hfill{(24.iv)}
\vskip.5pc
$\qquad\quad=\left(ip_0\frac{\partial}{\partial p_j}-\frac{[{\bf S}\land{\bf p}]_j}{\mu+p_0}\right)\Upsilon\SREV_{mn}=
{\textsf k}_j{\hat S}_{mn}$.
\vskip.8pc
Now, since each component projective representation $U^{(m)}:{\mathcal P}_+^\uparrow\to{\mathcal H}^{(m)})$
is irreducible, (24) imply that each ${\hat S}_{mn}$ is a multiple of the identity, so that
${\hat S}=\left[\begin{matrix}c_{11}& c_{12} \cr c_{21} & c_{22}\end{matrix}\right]$.
According to sect. 2, the further constraint $\SREV^2=\Id$ can be imposed; it is satisfied if and only if ${\hat S}^2=\Id$;
this implies that $\hat S={\hat S}^{-1}={\hat S}^\ast$ is a constant hermitean matrix with eigenvalues
$+1$ and $-1$, where ${\hat S}^\ast$ denotes the adjoint of $\hat S$; therefore
$$
\SREV=\Upsilon\,{\bf n}\cdot{\boldsymbol\sigma},\;\hbox{ where } {\bf n}\in\RR^3,\;\Vert{\bf n}\Vert=1\hbox{ and }
{\boldsymbol\sigma}=\left(\left[\begin{matrix}0&1\cr 1&0\end{matrix}\right],\left[\begin{matrix}0&-i\cr i&0\end{matrix}\right],
\left[\begin{matrix}1&0\cr 0&-1\end{matrix}\right]\right).
\eqno(25)
$$
If $\hat W=\left[\begin{matrix}w_{11}&w_{12}\cr w_{21}& w_{22}\end{matrix}\right]$ is any unitary $2\times 2$ matrix, then
$[\hat W,P_0]=[\hat W,P_j]=[\hat W,J_k]=[\hat W,K_j]=\nop$; such a matrix always exists such that
$\hat W\SREV\hat W^{-1}=\Upsilon\hat W{\bf n}\cdot{\boldsymbol\sigma}\hat W^{-1}=
\Upsilon\left[\begin{matrix}0& 1\cr 1& 0\end{matrix}\right]$.
The quantum theory of the system can be equivalently formulated by converting every operator $B$ into $\hat WB\hat W^{-1}$;
in so doing the operators
$P_0$, $P_j$, $J_k$ and $K_j$ remain unaltered because each of them has the form $\left[\begin{matrix}A&0\cr 0&A\end{matrix}\right]$,
while
$$\SREV=\Upsilon\left[\begin{matrix}0&1\cr 1&0\end{matrix}\right]\,.
\eqno(26)
$$
\subsubsection{Time reversal operator $\TREV$}
No we identify the time reversal operator $\TREV$ that completes the generalized projective representation of $\mathcal P$ that
extends the reducible projective representation
$U:{\mathcal P}_+^\uparrow\to{\mathcal U}(L_2(\RR^3,\CC^{2s+1},d\nu)$.
The conditions (8) imply
$$
\TREV_{mn}p_0=p_0\TREV_{mn}\,,\; \TREV_{mn}p_j=-p_j\TREV_{mn}\,,\; \TREV_{mn}{\textsf j}_k=
-{\textsf j}_k\TREV_{mn}\,,\; \TREV_{mn}{\textsf k}_j={\textsf k}_j\TREV_{mn}.
\eqno(27)
$$
The anti-unitary operator ${\mathcal K}$
satisfies the following relation
$$
{\mathcal K}p_j=p_j{\mathcal K},\quad  {\mathcal K}\frac{\partial}{\partial p_j}=\frac{\partial}{\partial p_j}{\mathcal K},\quad
{\mathcal K}\Upsilon=\Upsilon{\mathcal K},\quad{\mathcal K}^2=\Id\,.\eqno(28)
$$
Let us introduce the operator $\hat T=\left[\begin{matrix}{\hat T}_{11}& {\hat T}_{12}\cr {\hat T}_{21}&{\hat T}_{22}\end{matrix}\right]$,
with ${\hat T}_{mn}=\tau{\mathcal K}\Upsilon\TREV_{mn}$, that is unitary,
so that
$\TREV=\Upsilon{\mathcal K}\tau^{-1}\hat T\equiv\tau{\mathcal K}\Upsilon\hat T$. Relations (27), (23), (28) imply
\vskip.8pc\noindent
$\hat T_{mn} p_0=\tau{\mathcal K}\Upsilon\TREV_{mn}p_0=
\tau{\mathcal K}\Upsilon p_0\TREV_{mn}=\tau p_0{\mathcal K}\Upsilon \TREV_{mn}=
p_0\tau{\mathcal K}\Upsilon \TREV_{mn}=p_0\hat T_{mn}$,
\hfill{(29.i)}
\vskip.7pc\noindent
$\hat T_{mn} p_j=\tau{\mathcal K}\Upsilon\TREV_{mn}p_j=-\tau {\mathcal K}\Upsilon p_j \TREV_{mn}=
\tau p_j{\mathcal K}\Upsilon \TREV_{mn}=p_j\tau{\mathcal K}\Upsilon \TREV_{mn}  p_j\hat T_{mn}$,
\hfill{(29.ii)}
\vskip.7pc\noindent
$\hat T_{mn} {\textsf j}_k=\tau{\mathcal K}\Upsilon\TREV_{mn}{\textsf j}_k=
-\tau{\mathcal K}\Upsilon {\textsf j}_k\TREV_{mn}= $\hfill{(29.iii)}\par\noindent
$\qquad\quad -\tau{\mathcal K}\Upsilon\left(ip_l\frac{\partial}{\partial p_j}-ip_j\frac{\partial}{\partial p_l}\right)\TREV_{mn}
-\tau{\mathcal K}\Upsilon S_k\TREV_{mn}=$\par\noindent
$\qquad\quad\tau\left(ip_l\frac{\partial}{\partial p_j}-ip_j\frac{\partial}{\partial p_l}\right){\mathcal K}\Upsilon\TREV_{mn}
-\tau\overline{S_k}\tau^{-1}\tau{\mathcal K}\Upsilon\TREV_{mn}=$
\vskip.5pc
$\qquad\quad=\left(ip_l\frac{\partial}{\partial p_j}-ip_j\frac{\partial}{\partial p_l}\right)\tau{\mathcal K}\Upsilon  \TREV_{mn}+ S_k\tau{\mathcal K}\Upsilon\TREV_{mn}
=  {\textsf j}_k\hat T_{mn}$,
\vskip.7pc\noindent
$\hat T_{mn}{\textsf k}_j=\tau{\mathcal K}\Upsilon \TREV_{mn}{\textsf k}_j=\tau{\mathcal K}\Upsilon {\textsf k}_j \TREV_{mn}=
\tau{\mathcal K}\Upsilon\left(ip_0\frac{\partial}{\partial p_j}-\frac{[{\bf S}\land{\bf p}]_j}{\mu+p_0}\right) \TREV_{mn}=$ \hfill{(29.iv)}
\vskip.5pc $\qquad\quad=\tau\left(ip_0\frac{\partial}{\partial p_j}+\frac{[\overline{\bf S}\land{\bf p}]_j}{\mu+p_0}\right){\mathcal K}\Upsilon\hat \TREV_{mn}=
\left(ip_0\frac{\partial}{\partial p_j}+\frac{[\tau\overline{\bf S}\tau^{-1}\land{\bf p}]_j}{\mu+p_0}\right)\tau{\mathcal K}\Upsilon\hat \TREV_{mn}=
{\textsf k}_j\hat T_{mn}$.
\vskip.8pc\noindent
The irreducibility of each component $U^{(m)}:{\mathcal P}_+^\uparrow\to{\mathcal U}(L_2(\RR^3,\CC^{2s+1},d\nu))$
implies that ${\hat T}=\left[\begin{matrix}d_{11}& d_{12} \cr d_{21} & d_{22}\end{matrix}\right]$, with $d_{mn}$ constant.
\vskip.8pc
Further constraints are imposed by the condition
$\TREV^2=c$, with $c=\pm 1$. Now we have
$\TREV^2=\tau{\mathcal K}\Upsilon\hat T \tau{\mathcal K}\Upsilon\hat T=\tau\overline{\tau}\overline{\hat T}\hat T$;
therefore  $\TREV^2=\overline{\hat T}\hat T$ if $s\in\NN$ and $\TREV^2=-\overline{\hat T}\hat T$ if $s\in(\NN+\frac{1}{2})$.
It is clear that
there are always unitary constant matrices ${\hat T}=\left[\begin{matrix}d_{11}& d_{12} \cr d_{21} & d_{22}\end{matrix}\right]$ for which
$\TREV=(\tau{\mathcal K}\Upsilon\hat T)^2=\pm\Id$: it is sufficient that $\overline{\hat T}\hat T=\pm 1$; a trivial solution is ${\hat T}=\left[\begin{matrix}1& 0 \cr 0 & 1\end{matrix}\right]$,
that satisfies also (9);
but other less trivial solutions can easily singled out, such as $\hat T=\left[\begin{matrix}0& 1 \cr -1 & 0\end{matrix}\right]$.\par\noindent
So, extensions of $U^{(1)}\oplus U^{(2)}$ to generalized projective representations of the whole $\mathcal P$ are realized.
\vskip.5pc\noindent
{\bf Example 4.1.}
Let us study, for instance,
the case $s=0$, where $\tau=1$. The condition $\overline{\hat T}\hat T=\pm \Id$ entails ${\hat T}=c{\hat T}^t$, where $c=\pm 1$.
\vskip.4pc
If $c=1$, then ${\hat T}={\hat T}^t$, that implies $\hat T=\left[\begin{matrix}d_{11}& d_{12} \cr d_{21} & d_{22}\end{matrix}\right]=\left[\begin{matrix}d_{11}& d_{21} \cr d_{12} & d_{22}\end{matrix}\right]$, i.e. $d_{12}=d_{21}$.
Since ${\hat T}^2=\Id$ and $\hat T$ is unitary, $\hat T={\hat T}^{-1}={\hat T}^\ast$, i.e. $\hat T$ is hermitean and has two eigenvalues
$+1$ and $-1$. Therefore $\hat T={\bf n}\cdot \boldsymbol{\sigma}$, where $\bf n$ is a unit vector and ${\boldsymbol\sigma}$ is the three-operator
whose components are the Pauli matrices.
Condition (9) implies $\omega=\pm 1$, and if $\omega=1$ then $\hat T=\left[\begin{matrix}0&1\cr 1&0\end{matrix}\right]$,
whereas if $\omega=-1$ then $\hat T=\left[\begin{matrix}1&0\cr 0&-1\end{matrix}\right]$.
\par
If $c=-1$, then ${\hat T}=c{\hat T}^t$ implies $\hat T=\left[\begin{matrix}0& d \cr -d & 0\end{matrix}\right]$, with $d\in\CC$. In this case ${\hat T}^2=-1$,
so that $\left[\begin{matrix}-d^2&0 \cr 0&-d^2 \end{matrix}\right]=\left[\begin{matrix}-1& 0 \cr 0 & -1\end{matrix}\right]$, i.e $d=\pm 1$ and
we can take $\hat T=\left[\begin{matrix}0& 1\cr -1 & 0\end{matrix}\right]$.
\par
The required commutation relation (9) between $\SREV$ and $\TREV$ is satisfied in case $c=-1$ with $\omega=-1$. Indeed,
$$
\SREV\TREV=\Upsilon{\mathcal K}\Upsilon\left[\begin{matrix}0& 1 \cr 1 & 0\end{matrix}\right]\left[\begin{matrix}0& 1 \cr -1 & 0\end{matrix}\right]
={\mathcal K}\left[\begin{matrix}-1& 0 \cr 0 & 1\end{matrix}\right],
$$
and
$$
\TREV\SREV={\mathcal K}\left[\begin{matrix}0& 1 \cr -1 & 0\end{matrix}\right]\left[\begin{matrix}0& 1 \cr 1 & 0\end{matrix}\right]
={\mathcal K}\left[\begin{matrix}1& 0 \cr 0 & -1\end{matrix}\right]=-\SREV\TREV.
$$
\subsubsection{Irreducibility of the extension $U:{\mathcal P}\to{\mathcal U}({\mathcal H})$}
Now we show that, for each possible value of $s$,
there are irreducible generalized projective representation $U:{\mathcal P}\to{\mathcal V}({\mathcal H})$
that extend the reducible projective representations  $U:{\mathcal P}_+^\uparrow\to{\mathcal U}({\mathcal H})$
of the kind we are considering, that are irreducible.
\par
Let $A=\left[\begin{matrix}A_{11}& A_{12} \cr A_{21} & A_{22}\end{matrix}\right]$ be any self-adjoint operator of ${\mathcal H}=L_2(S_\mu^+,\CC^{2s+1},d\nu)\oplus L_2(S_\mu^+,\CC^{2s+1},d\nu)$, such that $[A,U_g]=\nop$ for all $g\in\mathcal P$, and therefore
$A$ commutes with all self-adjoint generators and with $\TREV$ and $\SREV$. From $[A,P_j]=\nop$ we imply that each $A_{mn}$ must be a function of $\bf p$:
$A_{mn}=a_{mn}({\bf p})$, and in particular $[A_{mn},p_j]=\nop$. Moreover, $[A, J_k]=\nop$ implies $[A_{mn},{\textsf j}_k]=\nop$.
Then, since $p_1,p_2,p_3,{\textsf j}_1,{\textsf j}_2,{\textsf j}_3$ are the generators of an irreducible projective representation of
$\mathcal E$ in the Hilbert space $L_2(S_\mu^+,\CC^{2s+1},d\nu)$, each $A_{mn}$ is a multiple of the identity: $A_{mn}=a_{mn}\Id$.
Therefore, the condition $[A,\SREV]=\nop$ implies $A=\left[\begin{matrix}a& b \cr b & a\end{matrix}\right]$.
In the generalized projective representation where $\hat T=\left[\begin{matrix}0& 1 \cr -1 & 0\end{matrix}\right]$
we have $\TREV=\tau^{-1}\Upsilon{\mathcal K}\left[\begin{matrix}0& 1 \cr -1 & 0\end{matrix}\right]$;
the condition $[A,\TREV]=\nop$ implies $b=0$. Therefore, a self-adjoint operator $A$ that commutes with all $U_g$, $g\in{\mathcal P}$
must have the form $A=\left[\begin{matrix}a& 0 \cr 0 & a\end{matrix}\right]\equiv a\Id$, and therefore the generalized projective representation $U$ is irreducible.
\subsection{The case $\sigma(\underline P)=S_\mu^+\cup S_\mu^-$}
Now we identify irreducible representations $U$ of $\mathcal P$ with $\sigma(\underline P)=S_\mu^+\cup S_\mu^-$
such that $U^+\mid_{{\mathcal P}_+^\uparrow}$, and hence $U^-\mid_{{\mathcal P}_+^\uparrow}$ by Prop. 3.1, is the direct sum of two irreducible projective representations $U^{(1)}$ and $U^{(2)}$
of ${\mathcal P}_+^\uparrow$.
\par
The aimed irreducibility of $U:{\mathcal P}\to{\mathcal V}({\mathcal H})$ forces its characterizing parameters  $\mu$ and $s$  to have the same values for
the reduced components $U^{(1)}$ and $U^{(2)}$;
hence, $U^{(1)}$ and $U^{(2)}$ must be unitarily isomorphic, so that they can be identified with two identical projective representations according to section 3.1.1.
\par
We consider the case where $s=0$, because its simplicity helps clearness. Each of the Hilbert spaces ${\mathcal M}^+$ of $U^{(1)}$ and ${\mathcal N}^+$ of
$U^{(2)}$ can be identified with $L_2(\RR^3,d\nu)$.
According to Prop. 3.1, both subspaces ${\mathcal M}={\mathcal M}^+\oplus {\mathcal M}^-$ and ${\mathcal N}={\mathcal N}^+\oplus {\mathcal N}^-$,
where ${\mathcal M}^-=\TREV{\mathcal M}^+$ and ${\mathcal N}^-=\TREV{\mathcal N}^+$ reduce $U\mid_{{\mathcal P}_+}$. Hence, every vector $\psi$ of the Hilbert space
$\mathcal H$ of the entire generalized projective representation of $\mathcal P$ can be uniquely decomposed as
$\psi=\psi_{{\mathcal M}^+}+\psi_{{\mathcal M}^-}+\psi_{{\mathcal N}^+}+\psi_{{\mathcal N}^-}$, with
$\psi_{{\mathcal M}^+}\in{\mathcal M}^+$, $\psi_{{\mathcal M}^-}\in{\mathcal M}^-$,
$\psi_{{\mathcal N}^+}\in{\mathcal N}^+$, $\psi_{{\mathcal N}^-}\in{\mathcal N}^-$, so that $\psi$ can be represented as a column vector
$\psi=\left[\begin{matrix}\psi_{{\mathcal M}^+}\cr\psi_{{\mathcal M}^-}\cr\psi_{{\mathcal N}^+}\cr\psi_{{\mathcal N}^-}\end{matrix}\right]$.
\par\noindent
In such a representation the self-adjoint generators of ${\mathcal P}_+^\uparrow$ are
$$
P_0=\left[\begin{matrix}p_0&0&0&0\cr 0&-p_0&0&0\cr 0&0&p_0&0\cr 0&0&0&-p_0\end{matrix}\right],\quad
P_j=\left[\begin{matrix}p_j&0&0&0\cr 0&p_j&0&0\cr 0&0&p_j&0\cr 0&0&0&p_j\end{matrix}\right],
$$
$$
J_k=\left[\begin{matrix}{\textsf j}_k&0&0&0\cr 0&{\textsf j}_k&0&0\cr 0&0&{\textsf j}_k&0\cr 0&0&0&{\textsf j}_k\end{matrix}\right],\quad
K_j=\left[\begin{matrix}{\textsf k}_j&0&0&0\cr 0&-{\textsf k}_j&0&0\cr 0&0&{\textsf k}_j&0\cr 0&0&0&-{\textsf k}_j\end{matrix}\right],
$$
According to Prop. 3.1, also the unitary operator $\TREV$ is reduced by $\mathcal M$ and $\mathcal N$, where its irreducible components,
by (21), are both $\left[\begin{matrix}0&1\cr 1&0\end{matrix}\right]$.
Then we have $\TREV=\left[\begin{matrix}0&1&0&0\cr 1&0&0&0\cr 0&0&0&1\cr 0&0&1&0\end{matrix}\right]$.
\vskip.5pc\noindent
In the case $\SREV^2=\Id$, since $\SREV$ is anti-unitary, by making use of (6) we find
$\SREV={\mathcal K}\left[\begin{matrix}0&s_1&0&s_2\cr s_1&0&s_2&0\cr 0&s_2&0&s_3\cr s_2&0&s_3&_0\end{matrix}\right]$.
\par
Now, let $A=\left[\begin{matrix}A_{11}&A_{12}&A_{13}&A_{14}\cr A_{21}&A_{22}&A_{23}&A_{24}\cr A_{31}&A_{32}&A_{33}&A_{34}\cr A_{41}&A_{42}&A_{43}&A_{44}
\end{matrix}\right]$ be any self-adjoint operator of $\mathcal H$;
the conditions $[A,P_0]=[A,P_j]=[A,J_k]=[A,K_j]=[A,\TREV]=[A,\SREV]=\nop$ are satisfied if and only if
$A=\left[\begin{matrix}a&0&b&0\cr 0&a&0&b\cr \overline b&0&c&0\cr 0&\overline b&0&c
\end{matrix}\right]$ where $a,c\in\RR$ and $b\in\CC$, provided that $a+\overline b=b+c$.
Therefore, there are self-adjoint operators $A$ that commute with
all $U_g\in U({\mathcal P})$, different from a multiple of the identity. We have to conclude that if $\SREV^2=1$ then $U:{\mathcal P}\to{\mathcal V}({\mathcal H})$ is reducible.
\par
Let us consider the case that $\SREV^2=-\Id$. We find that the conditions (6), (9) are satisfied if and only if
$\SREV={\mathcal K}\left[\begin{matrix}0&0&0&1\cr 0&0&1&0\cr 0&-1&0&0\cr -1&0&0&0\end{matrix}\right]$.
If $A$ is any self-adjoint operator of $\mathcal H$, then this time the conditions
$[A,P_0]=[A,P_j]=[A,J_k]=[A,K_j]=[A,\TREV]=[A,\SREV]=\nop$ imply
$A=\left[\begin{matrix}a&0&0&0\cr 0&a&0&0\cr 0&0&a&0\cr 0&0&0&a
\end{matrix}\right]\equiv a\Id$ with $a\in\RR$. Thus $U$ is irreducible.
\subsection{Resuming scheme}
Now we present a rational scheme of the re-determination attained in sections 3 and 4,
that classifies and identifies the irreducible generalized projective representations
of $\mathcal P$ with $\mu>0$.
\vskip.8pc
Let us denote the class of all irreducible generalized projective representations of $\mathcal P$ by ${\mathcal I}_{\mathcal P}$
(unitarily equivalent representations are identified in ${\mathcal I}_{\mathcal P}$). Then we have a first classification according to the characterizing parameters $\mu$ and $s$:
\vskip.5pc\noindent
{\bf C.1.}\quad ${\mathcal I}_{\mathcal P}=\cup_{\mu>0, s\in\frac{1}{2}\NN}\;{\mathcal I}_{\mathcal P}(\mu,s)$,
\vskip.5pc\noindent
where ${\mathcal I}_{\mathcal P}(\mu,s)$ is the class of all representations in
${\mathcal I}_{\mathcal P}$ such that $P_0^2-{\bf P}^2=\mu^2$ and $W^2=\mu s(s+1)$.
In its turn, each class ${\mathcal I}_{\mathcal P}(\mu,s)$ in {\bf C.1} can be decomposed as
\vskip.5pc\noindent
{\bf C.2.}\quad ${\mathcal I}_{\mathcal P}(\mu,s)=
{\mathcal I}_{\mathcal P}(S_\mu^+,s)\cup {\mathcal I}_{\mathcal P}(S_\mu^-,s)\cup{\mathcal I}_{\mathcal P}(S_\mu^+\cup S_\mu^-,s)$,
\vskip.5pc\noindent
where ${\mathcal I}_{\mathcal P}(S_\mu^\pm,s)$ is the class of all representations in ${\mathcal I}_{\mathcal P}(\mu,s)$ such that
$\sigma(\underline P)=S_\mu^\pm$, and
${\mathcal I}_{\mathcal P}(S_\mu^+\cup S_\mu^-,s)$ is the class of all representations in
${\mathcal I}_{\mathcal P}(\mu,s)$ such that $\sigma(\underline P)=S_\mu^+\cup S_\mu^-$.
Each component of ${\mathcal I}_{\mathcal P}(\mu,s)$ in {\bf C.2} can be further decomposed
into two sub-classes according to the reducibility of $U^+\mid_{{\mathcal P}_+^\uparrow}$ or $U^-\mid_{{\mathcal P}_+^\uparrow}$:
\vskip.5pc\noindent
{\bf C.3.u.}\quad ${\mathcal I}_{\mathcal P}(S_\mu^+,s)=
{\mathcal I}_{\mathcal P}(S_\mu^+,s,U^+{\rm irred.})\cup {\mathcal I}_{\mathcal P}(S_\mu^+,s,U^+{\rm red.})$,
\vskip.5pc\noindent
{\bf C.3.d.}\quad ${\mathcal I}_{\mathcal P}(S_\mu^-,s)=
{\mathcal I}_{\mathcal P}(S_\mu^-,s,U^-{\rm irred.})\cup {\mathcal I}_{\mathcal P}(S_\mu^-,s,U^-{\rm red.})$,
\vskip.5pc\noindent
{\bf C.3.s.}\quad ${\mathcal I}_{\mathcal P}(S_\mu^+\cup S_\mu^-,s)=
{\mathcal I}_{\mathcal P}(S_\mu^+\cup S_\mu^-,s,U^+{\rm irred.})\cup {\mathcal I}_{\mathcal P}(S_\mu^+\cup S_\mu^-,s,U^+{\rm red.})$,
\vskip.5pc\noindent
with obvious meaning of the notation.
\par
Finally, the components of the decompositions {\bf C.3} with
$U^+\mid_{{\mathcal P}_+^\uparrow}$ or $U^+\mid_{{\mathcal P}_+^\uparrow}$ irreducible are explicitly identified.
\vskip.5pc\noindent
{\bf C.u.}\quad
${\mathcal I}_{\mathcal P}(S_\mu^+,s,U^+{\rm irred.})$ contains a unique representation $U^{(u)}$, identified in section 3.1.1.
\vskip.5pc\noindent
{\bf C.d.}\quad
${\mathcal I}_{\mathcal P}(S_\mu^-,s,U^-{\rm irred.})$ contains a unique representation $U^{(d)}$, identified in section 3.1.2.
\begin{itemize}
\item[{\bf C.s.}]
${\mathcal I}_{\mathcal P}(S_\mu^+\cup S_\mu^-,s,U^+{\rm irred.})$\; contains six inequivalent representations $U^{(1)}$, $U^{(2)}$,...,$U^{(6)}$,
identified in section 3.2, all with the same Hilbert space ${\mathcal H}=L_2(\RR^3,\CC^{2s+1},d\nu)\oplus L_2(\RR^3,\CC^{2s+1},d\nu)$
and the same self-adjoint generators (20); they differ just for the different combinations of time reversal and space inversion operators.
\item[] $U^{(1)}$ has unitary $\TREV=\left[\begin{matrix}0&1\cr 1&0\end{matrix}\right]$ and unitary $\SREV=\Upsilon\left[\begin{matrix}1&0\cr 0&1\end{matrix}\right]$;
\item[] $U^{(2)}$ has unitary $\TREV=\left[\begin{matrix}0&1\cr 1&0\end{matrix}\right]$ and unitary $\SREV=\Upsilon\left[\begin{matrix}1&0\cr 0&-1\end{matrix}\right]$;
\item[] $U^{(3)}$ has unitary $\TREV=\left[\begin{matrix}0&1\cr 1&0\end{matrix}\right]$ and
anti-unitary $\SREV=\left[\begin{matrix}0&\tau\cr \tau&0\end{matrix}\right]{\mathcal K}$;
\item[] $U^{(4)}$ has unitary $\TREV=\left[\begin{matrix}0&1\cr 1&0\end{matrix}\right]$ and
anti-unitary $\SREV=\left[\begin{matrix}0&\tau\cr -\tau&0\end{matrix}\right]{\mathcal K}$;
\item[] $U^{(5)}$ has anti-unitary $\TREV=\tau{\mathcal K}\Upsilon\left[\begin{matrix}0&1\cr 1&0\end{matrix}\right]$ and anti-unitary$\SREV=\left[\begin{matrix}0&\tau\cr\tau&0\end{matrix}\right]{\mathcal K}$;
\item[] $U^{(6)}$ has anti-unitary $\TREV=\tau{\mathcal K}\Upsilon\left[\begin{matrix}0&1\cr 1&0\end{matrix}\right]$ and anti-unitary$\SREV=\left[\begin{matrix}0&\tau\cr-\tau&0\end{matrix}\right]{\mathcal K}$.
\end{itemize}
Hence, once fixed $\mu$ and $s$, there are eight inequivalent irreducible generalized projective representations
of $\mathcal P$ with $U^+\mid_{{\mathcal P}_+^\uparrow}$ or  $U^-\mid_{{\mathcal P}_+^\uparrow}$ irreducible.
The class of all such octets, for all permitted values of $\mu$ and $s$, does not exhaust ${\mathcal I}_{\mathcal P}$, because the components with $U^+\mid_{{\mathcal P}_+^\uparrow}$ or  $U^-\mid_{{\mathcal P}_+^\uparrow}$ reducible in the decompositions {\bf C.3} are not empty, as shown in sections 4.1 and 4.2.
\par
The whole class ${\mathcal I}_{\mathcal P}$ contains classes that are not considered in the literature about relativistic quantum theories of single particles;
for instance, in \cite{c13} only $U^{(u)}$, $U^{(d)}$, $U^{(1)}$ and $U^{(2)}$ are considered. Thus the present work identifies
two further (non-disjoint) robust classes of representations of $\mathcal P$ that should be considered for the formulation of relativistic quantum theories:
\vskip.5pc\noindent
${\mathcal I}_{\mathcal P}({\rm ant.}\SREV)$, i.e. the class that collects all representation of the kind $U^{(3)}$-$U^{(6)}$;
\vskip.5pc\noindent
${\mathcal I}_{\mathcal P}(U^\pm{\rm red.})$, i.e. the class of all representations in ${\mathcal I}_{\mathcal P}$ with $U^+\mid_{{\mathcal P}_+^\uparrow}$ or  $U^-\mid_{{\mathcal P}_+^\uparrow}$ reducible.
\section{Quantum theories of single particles}
In order to identify the specific theories of free particles, we interpret $\mathcal P$ as a group of changes of reference frame, according to special relativity. Hence,
given a reference frame $\Sigma$ in the class $\mathcal F$ of the (inertial) reference frames that move uniformly with respect to each other,
for every $g\in{\mathcal P}$ let $\Sigma_g$ denote the reference frame
related to $\Sigma$ just by $g$, and let $\textsf g:\RR^4\to\RR^4$ be the mapping such that if
$\underline x=(t,x_1,x_2,x_3)\equiv(x_0,{\bf x})$ is the vector of the time-space coordinates of an event with respect
to $\Sigma$, then $\textsf g(\underline x)$ is the vector of the time-space coordinates of that event with respect to $\Sigma_g$.
\par
Let us now consider an {\sl isolated} physical system. We formulate the following statement as a {\sl physical principle} valid for this system.
\begin{description}
\item[$\mathcal S${\it ym}]{\sl
All transformations of the Poincar\'e  group ${\mathcal P}$ are symmetry transformations for the system.}
\end{description}
An
implication of $\mathcal S${\it ym} is that for each symmetry transformation $g\in\mathcal P$ a specific {\sl quantum transformation}
$S_g:\Omega({\mathcal H})\to\Omega({\mathcal H})$, $A\to S_g[A]$ of the quantum observables exists,
we shall define
below through the following concept of
{\sl relative indistinguishability} between measuring procedures of quantum observables:
\begin{itemize}
\item[]
{\sl Given two reference frames $\Sigma_1$ and $\Sigma_2$ in $\mathcal F$, if a measuring procedure ${\mathcal M}_1$ is relatively to $\Sigma_1$ identical to what is another measuring procedure
${\mathcal M}_2$ relatively to $\Sigma_2$, we say that ${\mathcal M}_1$ and ${\mathcal M}_2$ are indistinguishable
relatively to $(\Sigma_1,\Sigma_2)$.}
\end{itemize}
Given $\Sigma_1$ and $\Sigma_2$ in $\mathcal F$, according to the the symmetry established by $\mathcal S${\it ym},
for every measuring procedure ${\mathcal M}_1$ another measuring procedure ${\mathcal M}_2$ must exist such that
${\mathcal M}_1$ and ${\mathcal M}_2$ are indistinguishable
relatively to $(\Sigma_1,\Sigma_2)$, otherwise symmetry does not hold.
\vskip.5pc\noindent
{\bf Definition 5.1.} {\sl Quantum transformation} (QT). \par\noindent
{\sl
Fixed any reference frame $\Sigma\in{\mathcal F}$,
for every $g\in{\mathcal P}$ the mapping
$$
S_g:\Omega({\mathcal H})\to \Omega({\mathcal H}),\quad A\to S_g[A]\,,\eqno(30)
$$
such that
the quantum observables $A$ and $S_g[A]$ are respectively measured by two measuring procedures
${\mathcal M}_1$ and ${\mathcal M}_2$ indistinguishable relatively to $(\Sigma,\Sigma_g)$,
is called {\sl Quantum Transformation} of $g$.}\vskip.5pc\noindent
The following properties are compelled by the present particular concept of quantum transformation.
\begin{itemize}
\item[(S.1)] {\sl Every $S_g:\Omega({\mathcal H})\to\Omega({\mathcal H})$ is bijective;}
\item[(S.2)] {\sl for every $A\in\Omega({\mathcal H})$ and every function $f$ such that  $f(A)\in\Omega({\mathcal H})$,
\item[] the equality $S_g[f(A)]=f(S_g[A])$ holds.}
\item[]
We show how this
property is compelled by conceptual coherence. Let us consider the two procedures ${\mathcal M}_1$ and ${\mathcal M}_2$
measuring $A$ and $S_g[A]$, that are indistinguishable relatively to $\Sigma$ and $\Sigma_g$, according to (QT).
General quantum theory \cite{v1} establishes that the quantum observable $B=f(A)$ can be measured by performing the same measuring
procedure ${\mathcal M}_1$ for measuring $A$ and then transforming the outcome $a$ of $A$ into the outcome $b=f(a)$ of
$f(A)$,
by the mathematical function $f$; the same argument applies to $S_g[A]$ and $D=f(S_g[A])$, of course.
Therefore, two measuring procedures that measure the quantum observables $f(A)$ and $f(S_g[A])$ can be realized
by transforming the outcomes yielded by the relatively indistinguishable ${\mathcal M}_1$ and ${\mathcal M}_2$ through the {\sl same} function $f$;
{\sl adding the application of the same mathematical function $f$ to the outcomes of ${\mathcal M}_1$
and ${\mathcal M}_2$ does not affect their relative indistinguishability}. Thus $S_g[f(A)]=f(S_g[A])$ follows.
\item[(S.3)] {\sl $S_{gh}=S_g\circ S_h$, for all $g,h\in\mathcal P$.}
\end{itemize}
Thus, from $(\mathcal S${\it ym}), by conceptual coherence, we have inferred the following further physical principle.
\begin{description}
\item[($\mathcal{QT}$)]{\sl
For every symmetry transformation $g\in\mathcal P$ a quantum transformation
$S_g:\Omega({\mathcal H})\to\Omega({\mathcal H})$ exists such that (S.1), (S.2) and (S.3) hold.}
\end{description}
Properties (S.1) and (S.2) were sufficient  \cite{N1} to prove that each quantum transformation $S_g$ is an {\sl automorphism} of the lattice $\Pi({\mathcal H})$
of the projection operators; therefore, according to Wigner theorem \cite{N1},\cite{b2}, a unitary or anti-unitary operator
${\tilde U}_{g}$ must exist such that
$$
S_{g}[A]={\tilde U}_gA{\tilde U}_g^{-1}\,,\quad\hbox{for every }A\in\Omega({\mathcal H}).
\eqno(31)
$$
Given any real function $\theta$ of $g$, the operators $\tilde U_g$ and $e^{i\theta(g)}\tilde U_g$
yield the same quantum transformation as $\tilde U_g$, i.e.
$\tilde U_gA \tilde U_g^{-1}= \left(e^{i\theta(g)}\tilde U_g\right)A\left(e^{i\theta(g)}\tilde U_g\right)^{-1}$,
and, hence, $U_g =e^{i\theta(g)}\tilde U_g$ can replace $\tilde U_g$ in the specific quantum theory of the system.
In particular, we can set $U_e=\Id$.
\vskip.5pc\noindent
{\bf Remark 5.1.}
It is important do not confuse the present concept of quantum transformation (QT) with the ``active'' concept more often adopted.
The transformation in this latter sense is obtained by ``Moving everything by an element [$g\in\mathcal P$]'' \cite{b3}.
The active concept, in fact, is not adequate for our approach. Let us explain why.
Let the apparatus $\mathcal M$ measuring $A$ be at rest with respect to $\Sigma$,
but with an ``internal'' component endowed with a velocity ${\bf v}$ with respect to $\Sigma$,
and let $g$ be a boost.
According to the {\sl active} concept, the apparatus ${\mathcal M}'$ measuring $S^{active}_g[A]$ is the apparatus $\mathcal M$
measuring $A$ endowed with the velocity $\bf u$ characterizing the boost $g$.
The apparatus ${\mathcal M}'$ is at rest with respect to $\Sigma_g$, of course, but the velocity of the moving component
is not $\bf v$ with respect to $\Sigma'$, because of the relativistic composition law of velocities. Therefore, $\mathcal M$ and ${\mathcal M}'$
are not indistinguishable with relation to $\Sigma$ and $\Sigma'$. Since such an indistinguishability is required in order that (S.2) holds,
the present approach could not be developed with the active concept of transformation.
\vskip.5pc
Condition (S.3) implies that $U_{g_1g_2}=\sigma(g_1,g_2)U_{g_1}U_{g_2}$ where $\sigma(g_1,g_2)$ is a complex number of modulus 1.
Hence, in general the correspondence $U:{\mathcal P}\to {\mathcal V}({\mathcal H})$, $g\to U_g$ realized according to these prescriptions is a
{\sl generalized} projective representation.
The properties of topological regularity of ${\mathcal P}_+^\uparrow$ (it is a connected Lie group!) can be translated into
the assumption that the correspondence $g\to S_g$ from ${\mathcal P}_+^\uparrow$ into the automorphisms of $\Pi({\mathcal H})$
is continuous, according to Bargmann topology \cite{N1}. Now,
it has been proved \cite{N1} that if the correspondence $g\to S_g$ assigning each transformation $g\in{\mathcal P}_+^\uparrow$ its quantum transformation
$S_g$ is continuous, then a choice
of the free phase $\theta(g)$ exists such that the restriction $U:{\mathcal P}_+^\uparrow\to {\mathcal U}({\mathcal H})$
turns out to be continuous, and therefore a continuous projective representation, according to Prop. 2.1.
\vskip.5pc
Thus, from the principles $\mathcal S${\it ym} and its``corollary'' ($\mathcal{QT}$), we have implied
that the Hilbert space of the quantum theory of an isolated system must necessarily be the Hilbert space of a generalized projective representation
of $\mathcal P$, that determine the quantum transformations as $S_g[A]=U_gAU_g^{-1}$; moreover, the restriction $U\mid_{{\mathcal P}_+^\uparrow}$
is continuous. As a consequence,
every theory of an isolated system can be constructed with the irreducible representations described by section 4.3.
In this section a more close identification of the specific theories of a {\sl
localizable free particle} is attained,
by introducing the further conditions that characterize such a
specific system.
In doing so we shall recover known results about the relativistic position operator, but
we find also that the further constraints imposed by localizability
require the further class of irreducible generalized projective representations identified in section 3.
\subsection{Localizable particle theories}
By {\sl localizable free particle}, shortly {\sl free particle}, we
mean an isolated system whose quantum theory is endowed with a unique {\sl position observable}, namely with a {\sl unique} triple
$(Q_1,Q_2,Q_3)\equiv{\bf Q}$ of self-adjoint operators,
whose components $Q_j$ are called coordinates, characterized by the following conditions.
\begin{itemize}
\item[({\it Q}.1)]
$[Q_j,Q_k]=\nop$, for all $j,k=1,2,3$.
\item[]
This condition establishes that a measurement of position yields
all three values of the coordinates of the same specimen of the particle.
\item[({\it Q}.2)]
The triple $(Q_1,Q_2,Q_3)\equiv{\bf Q}$ is characterized by
the specific properties of transformation of position with respect to the group $\mathcal P$,
i.e. by the specific mathematical relations between $\bf Q$ and $S_g[{\bf Q}]$.
\end{itemize}
{\bf Example 5.1.}
Let $\mathcal E$ be the Euclidean group, i.e. the group generated by spatial translations and rotations.
Condition ({\it Q}.2) implies that for $g\in\mathcal E\cup\{\trev,\srev\}$ the occurrence of ${\bf x}=(x_1,x_2,x_3)$ as outcome
of a measurement of the position at time $t=0$ in a reference frame
$\Sigma$ is equivalent to the occurrence of $\textsf g({\bf x})$ as outcome of a measurement of the position at time $t'=0$ in the reference frame $\Sigma_g$.
In formula,
\begin{itemize}
\item[({\it Q}.2]\hskip-3pt.a)\quad
$S_\trev[{\bf Q}]={\bf Q}$ and $S_\srev[{\bf Q}]=-{\bf Q}$, equivalent to $\TREV{\bf Q}={\bf Q}\TREV$ and $\SREV{\bf Q}=-{\bf Q}\SREV$,
\item[({\it Q}.2]\hskip-3pt.b)\quad
$S_g[{\bf Q}]=U_g{\bf Q}U_g^{-1}={\textsf g}({\bf Q})$ for every $g\in{\mathcal E}$.
\end{itemize}
So, the transformation property corresponding to a translation ${\textsf g}({\bf x})={\bf x}-{\bf a}$ is $U_g{\bf Q}U_g^{-1}={\bf Q}-{\bf a}\Id$.
It must be remarked that the extension of
(Q.2.b) cannot be extended to boosts is not available, as we shall explain in section 6.
\vskip.8pc
Conditions ({\it Q}.1) and ({\it Q}.2) imply relations that tie $\bf Q$ with the self-adjoint generators of $U$. For instance, if $g$ is a
translation by a length $a$ along $x_1$, so that $\textsf g({\bf x})=(x_1-a,x_2,x_3)$ and $U_g=e^{-iP_1a}$ hold, then ({\it Q}.2.b)
yields the transformation properties
$$
S_g[Q_k]=e^{-iP_1a}Q_ke^{iP_1a}=Q_k-\delta_{1k}a\;;
$$
by expanding with respect to $a$, these properties can be expressed as the commutation relation $[Q_k,P_1]=i\delta_{1k}$;
more generally, we imply the canonical commutation rules
$$
[Q_k,P_j]=i\delta_{jk}\;.\eqno(32.i)
$$
Analogously, the transformation properties with respect to spatial rotations imply
$$
[J_l,Q_j]=i\hat\epsilon_{ljk}Q_k. \eqno(32.ii)
$$
\par
In the quantum theory of a localizable free particle,
the system of operators $\{U({\mathcal P}),{\bf Q}\}$ can be reducible or not.
Following a customary habit, we refer to a particle for which $\{U({\mathcal P}),{\bf Q}\}$
is {\sl irreducible} as an {\sl elementary particle}.
\par
For elementary free particles, the generalized projective representation $U$ that realizes the quantum transformations
must be {\sl irreducible}. Let us explain why.
If $U$ were reducible, then a unitary operator $V$ would exist such that $[V,U_g]=\nop$ for all $g\in\mathcal P$,
but $[V,Q_j]\neq\nop$ for some $j$ (if $[V,Q_k]=\nop$ held for all $k$, then
$\{U({\mathcal P}),{\bf Q}\}$ would be reducible, and this is not possible for elementary particles).
Hence, if we define
$\hat Q_k=VQ_kV^{-1}$, then $\hat{\bf Q}\neq {\bf Q}$, while $VU_gV^{-1}=U_g$ for all $g\in\mathcal P$.
The mathematical relations between the operators $\hat{\bf Q}=V{\bf Q}V^{-1}$ and each $U_g=VU_gV^{-1}$ must be the same as the mathematical relations between
$\bf Q$ and that $U_g$, because
$\{U({\mathcal P}),{\bf Q}\}$ and $\{U({\mathcal P}),\hat{\bf Q}\}$ are unitarily isomorphic.
So, the triple $\hat{\bf Q}$ satisfies ({\it Q}.2), because $\bf Q$ does, and therefore
$\hat{\bf Q}$ would be a position operators in all respects.
Thus, for the same elementary particle two different position operators would exist, in contradiction with the required uniqueness.
\vskip.5pc
Accordingly,
by selecting the irreducible generalized projective representations $U$
of $\mathcal P$ that admit a triple $\bf Q$ satisfying ({\it Q}.1) and ({\it Q}.2) we identify the possible theories of single particle. Each theory so identified corresponds to a possible
kind of particle; theories that are unitarily {\sl inequivalent} correspond to different kind of particles. The actual existence in nature
of each of these particles is not matter that can be assessed in this theoretical paper.
\par
The work of sections 3 and 4 provided us with the structures an irreducible generalized projective representation $U$
of $\mathcal P$ can take, that extend
the family of representations taken into account in the literature; our selection will act on the enlarged domain
of representations with $U^\pm\mid_{{\mathcal P}_+^\uparrow}$ irreducible; the selection for $U^\pm\mid_{{\mathcal P}_+^\uparrow}$
reducible is matter of future work.
\par
In this section we show that if
the parameter $s$ that characterizes the irreducible generalized projective representation of the quantum theory of an elementary free particle
is zero, then
there are precisely identified cases such that
conditions ({\it Q}.1) and ({\it Q}.2.a), ({\it Q}.2.b) turn out to be sufficient to completely determine the position operator; this means that in these cases the lack of an explicit formulation, analogous to({\it Q}.2.b), of the
transformation properties with respect to boosts is irrelevant.
For the irreducible generalized projective representation with $\sigma(\underline P)=S_\mu^+$ and $U^+\mid_{{\mathcal P}_+^\uparrow}$ irreducible, the result we find in section 5.1 below agrees
with that known in the literature. For the case $\sigma(\underline P)=S_\mu^+\cup S_\mu^-$, however,
we show  in section 5.2 that the irreducible generalized projective representations for which the
position operator is determined belong to the new class ${\mathcal I}_{\mathcal P}(U^\pm{\rm red.})$.
\par
In the case $s>0$ there are no free particle theories completely determined by ({\it Q}.1), ({\it Q}.2.a) and ({\it Q}.2.b),
because of the absence of explicit transformation properties of position with respect to boosts.
In section 6 we shall investigate these theories in relation with the work of Jordan and Mukunda who attempted to fill this lack
by assuming a particular form of such a transformation property.
\subsection{Elementary particle: cases $s=0$, $\sigma(\underline P)=S_\mu^\pm+$ and $U^\pm\mid_{{\mathcal P}_+^\uparrow}$ irreducible}
Let us define the self-adjoint operators $F_j=i\frac{\partial}{\partial p_j}-\frac{i}{2p_0^2}p_j$ of the Hilbert space
$L_2(\RR^3,\CC^{2s+1},d\nu)$, known as Newton and Wigner operators \cite{N-W}.
Since $[F_j,P_k]=i\delta_{jk}$ straightforwardly holds, by (32.i) we imply
$Q_j-F_j=f_j(\bf P)$, i.e. $\left((Q_j-F_j)\psi\right)({\bf p})=f_j({\bf p})\psi({\bf p})$, where $f_j({\bf p})\in\Omega(\CC^{2s+1})$ for all ${\bf p}\in\RR^3$.
On the other hand, since $[J_j,F_k]=i{\hat\epsilon}_{jkl}F_l$, by (32.ii) we imply
$$
[J_j,f_k({\bf P})]=i{\hat\epsilon}_{jkl}f_l({\bf P}).\eqno(33)
$$
In case $s=0$, we have $S_k=0$, $\Omega(\CC^{2s+1})=\Omega(\CC)\equiv\RR$,  and $\tau=1$. Then (33)
together with $[J_j,P_k]=i{\hat\epsilon}_{jkl}P_l$ implies $f_j({\bf P})=h(\vert{\bf p}\vert)p_j$; by redefining $h(\vert{\bf p}\vert)=f(p_0)$, with
$p_0=\sqrt{\mu^2+{\bf p}^2}$, we have
$$
{\bf f}({\bf P})=f(p_0){\bf p},\hbox{ where }f(p_0)\in\Omega(\CC)\equiv\RR.\eqno(34)
$$
Now,  $\TREV F_j=F_j\TREV$ straightforwardly holds and ({\it Q}.2.a) implies $\TREV Q_j=Q_j\TREV$, so that we obtain
$\TREV f_j({\bf p})=f_j({\bf p})\TREV$; from this equality, since $\TREV={\mathcal K}\Upsilon$, by (34) we derive
$$
\overline{f(p_0)}=-f(p_0).\eqno(35)
$$
Since $f(p_0)\in\RR$, (35) implies $f(p_0)=0$.
Therefore, ${\bf Q}={\bf F}$. The condition $S_\srev[{\bf Q}]=-{\bf Q}$, i.e. $\SREV{\bf Q}=-{\bf Q}\SREV$ turns out to be trivially satisfied.
\par
Thus, if $s=0$, there is a unique position operator $\bf Q={\bf F}$
and it is completely determined by (Q.1) and ({\it Q}.2.a),({\it Q}.2.b).
This result agrees with the well known derivations of the Newton and Wigner operators as position operators \cite{N-W},\cite{J80}.
\vskip.5pc\noindent
By a quite similar derivation the same result, ${\bf Q}={\bf F}$ is obtained for $\sigma(\underline P)=S_\mu^-$
\subsection{The case $s=0$, $\sigma(\underline P)=S_\mu^+\cup S_\mu^-$ and $U^\pm\mid_{{\mathcal P}_+^\uparrow}$ irreducible}
The irreducible generalized projective representations with  $\sigma(\underline P)=S_\mu^+\cup S_\mu^-$ and $U^+\mid_{{\mathcal P}_+^\uparrow}$ irreducible
are explicitly identified in section 3.2.3.
To identify the triple $\bf Q$ representing the three coordinates of position we introduce the difference ${\bf D}={\bf Q}-\hat{\bf F}$, where
$\hat{\bf F}={\bf F}\Id\equiv\left[\begin{matrix}{\bf F}&0\cr 0&{\bf F}\end{matrix}\right]$ is the Newton-Wigner operator in this representation.
\par
Now, $[\hat F_j,P_k]=i\delta_{jk}$ and $[J_j,\hat F_k]=i\epsilon_{jkl}\hat F_k$ hold; on the other hand
$[Q_j,P_k]=i\delta_{jk}$ and $[J_j,Q_k]=i\epsilon_{jkl}Q_k$ are conditions to be satisfied by $\bf Q$ according to (32);
therefore, the relations $[D_j,P_k]=i\delta_{jk}$ and $[J_j,D_k]=i\epsilon_{jkl}D_k$ must hold too.
Then we have
$$
{\bf Q}=\hat{\bf F}+{\bf D},\quad\hbox{where}\quad D_j=\left[\begin{matrix}d^{(j)}_{11}({\bf p})&d^{(j)}_{12}({\bf p})\cr d^{(j)}_{21}({\bf p})&d^{(j)}_{22}({\bf p})\end{matrix}\right],
\eqno(36)
$$
$$
\hbox{with the conditions }\;[{\textsf j}_j,d^{(k)}_{nm}({\bf p})]=i\hat\epsilon_{j,k,l}d^{(l)}_{nm}({\bf p}),\hbox{ for all }{\bf p}\in\RR^3\;,
\eqno(37)
$$
each $d^{(j)}_{mn}({\bf p})$ being a $(2s+1)\times(2s+1)$ matrix such that ${d^{(j)}_{mn}}^\ast({\bf p})=d^{(j)}_{nm}({\bf p})$.
\vskip.5pc
For $s=0$, similarly to what happens in the case $\sigma(\underline P)=S_\mu^+$,
(37) and $[{\textsf j}_j,p_k]=i\hat\epsilon_{jkl}p_l$ imply $d^{(j)}_{mn}({\bf p})=d_{mn}(p_0)p_j$, with $d_{mn}(p_0)\in\CC$.
Hence
$$
Q_j=F_j+D_j,\hbox{ where }D_j=\left[\begin{matrix}d_{11}(p_0)&d_{12}(p_0)\cr d_{21}(p_0)&d_{22}(p_0)\end{matrix}\right]p_j\;\hbox{ and }\; d^\ast_{mn}(p_0)=d_{nm}(p_0).\eqno(38)
$$
So, to determine $\bf Q$ we have to determine the functions $d_{mn}$ of $p_0$; the conditions $\TREV{\bf Q}={\bf Q}\TREV$ and $\SREV{\bf Q}=-{\bf Q}\SREV$
can help in solving the indeterminacy. However, according to section 3.2,
now the explicit form of $\TREV$ and $\SREV$ depend on their unitary or anti-unitary character; so we shall explore all different combinations.
\vskip.8pc\noindent
{\bf (UU)} Let us start with the case where $\TREV$ is unitary and $\SREV$ is unitary too, which is the only case considered in the literature.
According to (21),
$\TREV=\left[\begin{matrix}0&1\cr 1&0\end{matrix}\right]$, while $\SREV=\Upsilon\left[\begin{matrix}1&0\cr 0&1\end{matrix}\right]$
or $\SREV=\Upsilon\left[\begin{matrix}1&0\cr 0&-1\end{matrix}\right]$. By making use of this explicit form of $\TREV$ we find that
$$
\TREV{\bf Q}={\bf Q}\TREV\quad\hbox{implies}\quad
D_j=\left[\begin{matrix}d_1(p_0)&d_2(p_0)\cr d_2(p_0)&d_1(p_0)\end{matrix}\right]p_j.\eqno(39)
$$
i) If $\SREV=\Upsilon\left[\begin{matrix}1&0\cr 0&1\end{matrix}\right]$, then $\SREV{\bf Q}=-{\bf Q}\SREV$ is always satisfied.
Hence $\bf Q$ remains undetermined.
\vskip.5pc\noindent
ii) If $\SREV=\Upsilon\left[\begin{matrix}1&0\cr 0&-1\end{matrix}\right]$ then $\SREV{\bf Q}=-{\bf Q}\SREV$ holds whenever $d_2(p_0)=0$.
\par\noindent
{\sl Therefore, in the combination ({\bf UU}) the position operaqtor $\bf Q$ is undetermined.}
\vskip.8pc\noindent
{\bf (UA)} In the case that $\TREV$ is unitary and $\SREV$ is anti-unitary,
according to (21) and Prop.3.2, we have
$\TREV=\left[\begin{matrix}0&1\cr 1&0\end{matrix}\right]$, while $\SREV={\mathcal K}\left[\begin{matrix}0&1\cr 1&0\end{matrix}\right]$
or $\SREV={\mathcal K}\left[\begin{matrix}0&1\cr -1&0\end{matrix}\right]$.
\vskip.5pc\noindent
i) If $\SREV={\mathcal K}\left[\begin{matrix}0&1\cr -1&0\end{matrix}\right]$, then $\SREV{\bf Q}=-{\bf Q}\SREV$ is satisfied whenever
$d_1(p_0)=0$.
\par\noindent
Therefore $\bf Q$ is undetermined.
\vskip.5pc\noindent
ii) If $\SREV={\mathcal K}\left[\begin{matrix}0&1\cr 1&0\end{matrix}\right]$, then $\SREV{\bf Q}=-{\bf Q}\SREV$ implies $d_1(p_0)=d_2(p_0)=0$, i.e. $D_j=0$.
\par\noindent
{\sl Therefore $\bf Q$ is uniquely determined, and ${\bf Q}=\hat{\bf F}$.}
\vskip.8pc\noindent
{\bf (AA)} In the case that $\TREV$ is anti-unitary, $\SREV$ must be anti-unitary too, otherwise $\sigma(\underline P)=S_\mu^\pm$.
According to Prop. 3.2 and (21)
$\TREV={\mathcal K}\Upsilon\left[\begin{matrix}1&0\cr 0&1\end{matrix}\right]$, while $\SREV={\mathcal K}\left[\begin{matrix}0&1\cr 1&0\end{matrix}\right]$
or $\SREV={\mathcal K}\left[\begin{matrix}0&1\cr -1&0\end{matrix}\right]$.
With this explicit form of $\TREV$ we find that
$$
\TREV{\bf Q}={\bf Q}\TREV\quad\hbox{implies}\quad D_j=\left[\begin{matrix}0&id(p_0)\cr -id(p_0)&0)\end{matrix}\right]p_j.\eqno(40)
$$
i) If $\SREV={\mathcal K}\left[\begin{matrix}0&1\cr -1&0\end{matrix}\right]$ then $\SREV{\bf Q}=-{\bf Q}\SREV$ is always satisfied by (40).
\par\noindent
Therefore $\bf Q$ is undetermined.
\vskip.5pc\noindent
ii) If $\SREV={\mathcal K}\left[\begin{matrix}0&1\cr 1&0\end{matrix}\right]$ then $\SREV{\bf Q}=-{\bf Q}\SREV$ implies $d(p_0)=0$, i.e. $D_j=0$.
\par\noindent
{\sl Therefore $\bf Q$ is uniquely determined and ${\bf Q}=\hat{\bf F}$.}
\vskip.8pc
Thus, conditions ({\it Q}.1) and ({\it Q}.2.a), ({\it Q}.2.b) determine $\bf Q$ only in the cases ({\bf UA}.ii) and ({\bf UU}.ii), where
$\SREV$ is anti-unitary. This anti-unitarity, however, is perfectly consistent with the respective theories.
\subsection{Klein-Gordon particles}
For every value of the characterizing parameter $\mu>0$ four inequivalent theories of single particle have been singled out in sections 5.2 and 5.3,
according to the methodological commitments of the present approach, that prevent from the shortcomings of canonical
quantization.
In each of them the position operator is uniquely determined as the Newton-Wigner operator.
\par
To complete the theories, we derive the explicit form of the wave equations.
In all theories of the present approach
time evolution from time $0$ to time $t$ is a translation of time, operated by the unitary operator $e^{-iP_0t}$; therefore if the state vector is $\psi$
at time $0$, then it is $\psi_t=e^{-iP_0t}\psi$ at time $t$, so that Schroedinger equation
$$
i\frac{\partial}{\partial t}\psi_t=P_0\psi_t\,\eqno{(41)}
$$
immediately follows. The explicit wave equation is attained by replacing $P_0$ with the the specific time translation operator,
explicitly known in each specific theory.
\par
In order to explore how previous theories for spin 0 particles relate to the present ones,
we re-formulate these last in the following equivalent forms, obtained by means of unitary transformations
operated by the unitary operator $Z=Z_1Z_2$, where $Z_2=\frac{1}{\sqrt{p_0}}\Id$ and $Z_1$ is the inverse of the {\sl Fourier-Plancherel} operator.
\begin{itemize}
\item[$\mathcal T$.1]
The theory based on the irreducible representation identified by $\sigma(\underline P)=S_\mu^+$ and $s=0$, according to section 5.2,
can be equivalently reformulated in the Hilbert space
${\mathcal H}=Z\left(L_2(\RR^3,d\nu)\right)\equiv L_2(\RR^3)$.
Here the self-adjoint generators are ${\hat P}_j=-i\frac{\partial}{\partial x_j}$, ${\hat P}_0=\sqrt{\mu^2+{\nabla}^2}$,
${\hat J}_k=-i\left(x_l\frac{\partial}{\partial x_j}-x_j\frac{\partial}{\partial x_l}\right)$,
${\hat K}_j=\frac{1}{2}(x_j{\hat P}_0+{\hat P}_0x_j)$,
while ${\hat \SREV}=\Upsilon$, ${\hat \TREV}={\mathcal K}$. The Newton-Wigner operator representing position, in this representation becomes
the multiplication operator ${\hat Q}_j$, defined by ${\hat Q}_j\psi({\bf x})=x_j\psi({\bf x})$.
Accordingly, the wave equation is
$$
i\frac{\partial}{\partial t}\psi_t({\bf x})=\sqrt{\mu^2-\nabla^2}\psi_t({\bf x})\,.\eqno{(42)}
$$
\item[$\mathcal T$.2]
The new formulation of
the theory based on the irreducible representation identified by $\sigma(\underline P)=S_\mu^-$ and $s=0$, according to section 5.2,
differs from $\mathcal T$.1 just for the time translation generator, which now is ${\hat P}_0=-\sqrt{\mu^2+{\nabla}^2}$,
and hence the wave equation is
$$
i\frac{\partial}{\partial t}\psi_t({\bf x})=-\sqrt{\mu^2-\nabla^2}\psi_t({\bf x}).\eqno{(43)}
$$
\item[$\mathcal T$.3]
The theory corresponding to ({\bf UA}.ii) in section 5.3, that is based on the irreducible representation of section 3.2 with $\sigma(\underline P)=S_\mu^+\cup S_\mu^-$ and $s=0$,
identified by $\TREV=\left[\begin{matrix}0&1\cr 1&0\end{matrix}\right]$
and
$\SREV={\mathcal K}\left[\begin{matrix}0&1\cr 1&0\end{matrix}\right]$, can be equivalently reformulated in the Hilbert space
${\mathcal H}=Z\left(L_2(\RR^3,d\nu)\oplus L_2(\RR^3,d\nu)\right)\equiv L_2(\RR^3)\oplus L_2(\RR^3)$; the new self-adjoint generators are
${\hat P}_j=\left[\begin{matrix}-i\frac{\partial}{\partial x_j}&0\cr 0&-i\frac{\partial}{\partial x_j}\end{matrix}\right]$,
${\hat P}_0=\sqrt{\mu^2-\nabla^2}\left[\begin{matrix}1&0\cr 0&-1\end{matrix}\right]$,
${\hat J}_k=-i\left(x_l\frac{\partial}{\partial x_j}-x_j\frac{\partial}{\partial x_l}\right)\left[\begin{matrix}1&0\cr 0& 1\end{matrix}\right]$;
${\hat K}_j=\frac{1}{2}\left(x_j\sqrt{\mu^2-\nabla^2}+\sqrt{\mu^2-\nabla^2}x_j\right)\left[\begin{matrix}1&0\cr 0&-1\end{matrix}\right]$.
The position operator is
${\hat Q}_j=\left[\begin{matrix}x_j&0\cr 0&x_j\end{matrix}\right]$.
\par\noindent
The wave equation is
$$
i\frac{\partial}{\partial t}\left[\begin{matrix}\psi^+_t({\bf x})\cr \psi^-_t({\bf x})\end{matrix}\right]
=\left[\begin{matrix}\sqrt{\mu^2-\nabla^2}\psi^+_t({\bf x})\cr -\sqrt{\mu^2-\nabla^2}\psi^-_t({\bf x})\end{matrix}\right]\,.\eqno(44)
$$
\item[$\mathcal T$.4]
The theory corresponding to ({\bf AA}.ii) in section 5.3, based on the irreducible representation of section 3.2 with $\sigma(\underline P)=S_\mu^+\cup S_\mu^-$ and $s=0$,
identified by $\TREV=\Upsilon{\mathcal K}\left[\begin{matrix}1&0\cr 0&1\end{matrix}\right]$
and
$\SREV={\mathcal K}\left[\begin{matrix}0&1\cr 1&0\end{matrix}\right]$,
differs from $\mathcal T$.3 only for these operators.
\end{itemize}
\vskip.5pc\noindent
The early theory for spin 0 particle establishes that the wave equation is Klein-Gordon equation
$$
\left(\frac{\partial^2}{\partial t^2}-\nabla^2\right)\psi_t({\bf x})=
-m^2\psi_t({\bf x})\,,\eqno{(45)}
$$
that is second order with respect to time.
This is the first evident difference with respect to theories $\mathcal T$.1-$\mathcal T$.4,
where all wave equations are first order.
However, if in each theory $\mathcal T$.1-$\mathcal T$.4 the respective wave equation is solved by $\psi_t$, then the derivative of the
equation with respect to time yields
$-\frac{\partial^2}{\partial t^2}\psi_t=iP_0\frac{\partial}{\partial t}\psi_t=P_0^2\psi_t$,
since $\frac{\partial}{\partial t}$ commutes with $P_0$ in all cases, obtaining
$$
\frac{\partial^2}{\partial t^2}\psi_t({\bf x})-{\nabla}^2\psi_t({\bf x})=-\mu^2\psi_t({\bf x})\,.\eqno(46)
$$
Hence, in all theories $\mathcal T$.1-$\mathcal T$.4 equation (46) is implied, which coincides with
Klein-Gordon equation,
once identified $\mu$ with the mass $m$. However, this coincidence does not mean that theories $\mathcal T$.1-$\mathcal T$.4
are equivalent to Klein-Gordon theory.
A first difference is that according to our approach there are {\sl four} inequivalent theories for spin 0 and ``mass'' $\mu$ particles.
In $\mathcal T$.1 there are no wave functions corresponding to negative spectral values of $P_0$. In $\mathcal T$.2 the positive values are
forbidden. Klein-Gordon theory does not exhibit this differentiation. In particular, the space of the vector states is only one,
namely the space generated by the solutions of (45).
\par
A second obvious evidence of non-equivalence is the difference between the set of solutions of the
respective wave equations:
while all solutions of the wave equations of $\mathcal T$.1-$\mathcal T$.4 are solution of Klein-Gordon equation,
the converse is not true, in general.
\vskip.5pc
A third important difference concerns with the physical interpretation and its consistency.
By means of mathematical manipulation it can be implied that for every solution $\psi_t$ of Klein-Gordon equation (45)
the following equation holds.
$$
\frac{\partial}{\partial t}\left\{\frac{i}{2\mu}\left(\overline{\psi_t}\frac{\partial}{\partial t}\psi_t-{\psi_t}\frac{\partial}{\partial t}\overline{\psi_t}
\right)\right\}
={\mathbf\nabla}\cdot\left\{\frac{i}{2\mu}\left(\psi_t\nabla\overline{\psi_t}-\overline{\psi_t}\nabla\psi_t\right)\right\}\,;\eqno(47)
$$
since it has the form of a continuity equation for a quantity whose density is
$\hat\rho(t,{\bf x})=\frac{i}{2\mu}\left(\overline{\psi_t}\frac{\partial}{\partial t}\psi_t-{\psi_t}\frac{\partial}{\partial t}\overline{\psi_t}\right)$
and whose current density is $\hat{\bf j}(t,{\bf x})=\frac{i}{2\mu}\left(\psi_t\nabla\overline{\psi_t}-\overline{\psi_t}\nabla\psi_t\right)$,
in Klein-Gordon theory $\hat\rho$ was interpreted as the {\sl density probability of position}, and $\hat{\bf j}$ as
the {\sl current density of the position probability}.
This interpretation is the source of serious problems for Klein-Gordon theory.
According to it, indeed, the knowledge of the quantum state vector $\psi_t$ at a given time $t$ would be not sufficient to determine the
probability density of position at that time, because $\hat\rho$ requires also the time derivative of $\psi_t$.
This fact directly conflicts with the general laws of quantum theory, where the probability density
of {\sl any} quantum observable $A$ at time $t$ is determined by the quantum state $\psi_t$
and no derivative is involved; indeed,
according to general quantum theory \cite{v1} the probability
that $A$ has a value in the interval $(a,b]$ is $p(\Delta)=\langle\psi_t\mid (E_b-E_a)\psi_t\rangle$, where $E$ is
the resolution of the identity of the self-adjoint operator $A$; the density probability is just the Radon-Nicodym
derivative of this probability measure $p$.
\par
A way to overcome the difficulty was proposed by Feshbach and Villars \cite{FV}.
They derive an equivalent form of Klein-Gordon equation as a first order equation
$i\frac{\partial}{\partial t}\Psi_t=H\Psi_t$
for the state vector $\Psi_t=\left[\begin{matrix}\phi_t\cr \chi_t\end{matrix}\right]$, where
$\phi_t=\frac{1}{\sqrt{2}}(\psi_t+\frac{1}{m}\frac{\partial}{\partial t}\psi_t)$,
$\chi_t=\frac{1}{\sqrt{2}}(\psi_t-\frac{1}{m}\frac{\partial}{\partial t}\psi_t)$, and
$H=(\sigma_3+\sigma_2)\frac{1}{2m}(\nabla+m\sigma_3)$;
in this representation $\hat\rho=\vert\phi_t\vert^2-\vert\chi_t\vert^2$, that is determined only by the new quantum state $\Psi_t$.
The proposal of Feshbach and Villars requires to drastically change the interpretation of Klein-Gordon equation:
the quantum state is not $\psi_t$, but $\Psi_t$. The minus sign in $\hat\rho$ forbids to interpret it as probability density of position;
Feshbach and Villars proposed to interpret it as density probability of {\sl charge}.
Nevertheless, the acceptance of Feshbach and Villars proposal requires another consistency test for $\hat\rho$,
that is to say covariance with respect to boosts, that implies, according to Barut and Malin \cite{BM68}, that
$\hat\rho$ must be the time component of a four-vector. Barut and Malin proved that is not the case.
\vskip.5pc
The theories $\mathcal T$.1-$\mathcal T$.4 of the present approach are not affected by these problems.
In all of them the position is represented by the multiplication operator, and therefore
the probability density of position must necessarily be $\rho(t,{\bf x})=\vert\psi_t({\bf x})\vert^2$.
Therefore the quantum state at a given time determines the probability density of position.
\par
The covariance properties with respect to boosts, according to ({\it Q}.2), are explicitly expressed by
$S_g[{\bf Q}]=e^{iK_j\varphi(u)}{\bf Q}e^{-iK_j\varphi(u)}$, being $K_j$ and $\bf Q$ explicitly known,
and there is no need of a four-density concept.
On the other hand, the existence of
a four-current density $(\hat\rho,\hat{\bf j})$ probability
satisfying the continuity equation and that transforms
as a four-vector is not compelled in the present approach. Indeed, it
has not a formal proof; it could be assumed to hold by {\sl analogy} with the electrical four-current
density in non-quantum relativistic theory; but the proof in this last theory is based on the existence of a {\sl real motion}
of the charges characterized by a {\sl real velocity}; then it turns out to be clear that such a proof cannot be repeated
for the probability density in quantum case.
\vskip.5pc\noindent
{\bf Remark 5.2.}
The further problems plaguing Klein-Gordon theory, e.g. those connected with transitions to lower levels,
arise if the particle is allowed to interact. Therefore they can be considered in the present approach only once the theory of interacting particle is developed
according to our methodological commitments.
\section{Transformation properties relative to boosts}
Apart from the cases identified in sections 5.2 and 5.3, in general
conditions ({\it Q}.1) and ({\it Q}.2.a,b)
do not univocally determine the position operator $\bf Q$, in particular when $s>0$;
e.g., for the case $\sigma(\underline P)=S_\mu^+$ and $U^+\mid_{{\mathcal P}_+^\uparrow}$ irreducible, Jordan concludes that
``For nonzero spin [i.e. $s>0$] there is more than one Hermitian operator [$\bf Q$] that transforms as a position operator should for translations, rotations,
and time reversal"  \cite{J80}.
A cause of such an indeterminacy is that
conditions ({\it Q}.2.a), ({\it Q}.2.b) miss the explicit mathematical relations that express
the transformation properties of $\bf Q$ with respect to boosts.
It is obvious that these further properties would impose further constraints, that could contribute to
solve the indeterminacy of $\bf Q$.
\par
In fact, for $s>0$ the extension of ({\it Q}.2.b) to boosts is not an easy matter.
To realize such an extension we have to consider two reference frames $\Sigma$ and $\Sigma_g$ where $g$ is a boost,
for instance the boost characterized by a relative velocity ${\bf u}=(u,0,0)$.
The explicit determination of the relation between $S_g[{\bf Q}]$ and $\bf Q$ requires,
given the position outcome $\bf x$ at time $t=0$ with respect to $\Sigma$, to identify the corresponding position outcome ${\bf y}_{\bf x}$
with respect to $\Sigma_g$, but at time $t'=0$ with respect to $\Sigma_g$, because of the condition of relative
indistinguishability dictated by the concept of quantum transformation (QT).
Special relativity does not provide such a correspondence
as a functional relation like those in ({\it Q}.2.b);
in fact, if the outcome of position is ${\bf x}=(x_1,x_2,x_3)$ at time $t=0$ in $\Sigma$, then
according to special relativity
we can state only that it corresponds to the position ${\bf y}=\left(\frac{x_1}{\sqrt{1-u^2}},x_2,x_3\right)$ in $\Sigma_g$, but at the time
$t'=\frac{-ux_1}{\sqrt{1-u^2}}$, not at $t'=0$!
\par
In standard presentations, based on canonical quantization, it is assumed that covariance of position under boosts implies that
for every quantum state a {\sl probability current density} $\hat{\bf j}(t,{\bf x})$ must exist, which satisfies the continuity equation
$\frac{\partial}{\partial t}\hat\rho={\bf\nabla}\cdot \hat{\bf j}$, where $\hat\rho(t,{\bf x})$ is the probability density of position,
such that $\hat{\underline j}=(\hat\rho,\hat{\bf j})$ transforms as a four-vector.
However, as already noticed in section 5.4, an explicit proof of this implication does not exist, not even in the present approach; therefore,
such assumption is ruled out by the methodological commitments of the present work.
\par
A way to escape the problem could be to introduce a {\sl time-space position} $\underline Q=(Q_0,{\bf Q})$, where the operator
$Q_0$ is just {\sl time} quantum observable, that represents the time when the measurement of the spatial coordinates represented by $\bf Q$ occurs.
Then, according to special relativity we could set
$S_g[Q_0]=\frac{Q_0-uQ_1}{\sqrt{1-u^2}}$, $S_g[Q_1]=\frac{Q_1-uQ_0}{\sqrt{1-u^2}}$, $S_g[Q_2]=Q_2$, $S_g[Q_3]=Q_3$.
But in the quantum theory of a localizable particle such a time cannot be a quantum observable.
Indeed, let us suppose that $Q_0$ is a quantum observable representing this time,
so that the four-operator $\underline Q=(Q_0,Q_1,Q_2,Q_3)$ represents the time-space coordinates of the particle.
Accordingly, $[Q_\alpha,Q_\beta]=\nop$, for all $\alpha,\beta\in\{0,1,2,3,\}$.
Furthermore, since for every time-space translation $g_\alpha$ such that $U_{g_\alpha}=e^{-iP_\alpha a}$
we have $S_{g_\alpha}[Q_\beta]=e^{-iP_\alpha a}Q_\beta e^{iP_\alpha a}=Q_\beta-\delta_{\alpha\beta}a$, also $[Q_\alpha,P_\beta]=i\delta_{\alpha\beta}$ holds;
in particular $[Q_0,P_0]=i$ and $[Q_0,P_j]=\nop$ should hold.
\par
Let us consider the case that $\sigma(\underline P)=S_\mu^+$ with $U^+\mid_{{\mathcal P}_+^\uparrow}$
irreducible. According to section 3.1.1,
$[Q_0,P_j]=\nop$ implies $\left(Q_0\psi\right)({\bf p})=q_0({\bf p})\psi({\bf p})$, where $q_0({\bf p})\in\Omega(\CC^{2s+1})$, for every ${\bf p}\in\RR^3$.
Therefore $P_0Q_0=p_0Q_0=Q_0p_0=Q_0P_0$, i.e. $[Q_0.P_0]=\nop$, in contradiction with $[Q_0,P_0]=i$.
In the alternative case $\sigma(\underline P)=S_\mu^+\cup S_\mu^-$ with $U^+\mid_{{\mathcal P}_+^\uparrow}$ irreducible, the same argument
leads to $[Q_0,P_0]=\left[\begin{matrix} 0&2q_{12}({\bf p})\cr 2q_{21}({\bf p})&0\end{matrix}\right]$,
where each $q_{mn}({\bf p})$ is a $(2s+1)\times(2s+1)$ matrix.
Also in this case the hypothesis that $Q_0$ is an observable contradicts $[Q_0,P_0]=i$.
\par
This proof can be easily extended to the irreducible generalized projective representations of section 4 for which $U^+\mid_{{\mathcal P}_+^\uparrow}$ is reducible.
\vskip.5pc
So, the hope of completing the development of the quantum theory of an elementary free particle remains tied to the possibility of attaining an
explicit mathematical relation expressing the transformation properties of $\bf Q$ with respect to boosts, that is to say, to the possibility of finding the extension
of (32) to $[K_j,Q_k]$.
In theories $\mathcal T$.1-$\mathcal T$.4, where ({\it Q}.1), ({\it Q}.2.a),({\it Q}.2.b) are sufficient to determine $\bf Q$,
the commutation relation $[K_j,Q_k]$ can be simply calculated, being $K_j$ and
$\bf Q$ explicitly known, and it turns out to be
$$
[K_j,Q_k]=\frac{1}{2}\left(Q_j[P_0,Q_k]+[P_0,Q_k]Q_j\right).\eqno(JM)
$$
Then the theory can be tentatively developed by assuming (JM) to hold in all cases, in particular when $s>0$.
It is evident that this extension is supported only by heuristic arguments.
In fact, in \cite{CJS} (JM) was derived by making use of canonical quantization, a method extraneous to the present
work; it turns out to be also the {\sl canonical quantization} of the transformation property derived in \cite{JMLL2}. However, the
consistency of its implications can be explored.
Jordan and Mukunda \cite{JM} checked the consistency of (JM) with the structural properties (1) of ${\mathcal P}_+^\uparrow$ and with the transformation properties
({\it Q}.2.b) of $\bf Q$ with respect to the Euclidean group $\mathcal E$.
The results attained by Jordan and Mukunda are not homogeneous;
an extract is given in section 6.1.
It is important to remark, however, that in their work \cite{JM} these authors assume the transformation properties of $\bf Q$ with respect to the subgroup ${\mathcal P}_+^\uparrow$,
ignoring the transformation properties with respect to time reversal and space inversion, i.e. $\TREV{\bf Q}={\bf Q}\TREV$ and $\SREV{\bf Q}=-{\bf Q}\SREV$
are not taken into account.
Sections 5.2 and 5.3 showed that in fact these conditions have a decisive role in determining the position operator;
therefore, the task of checking the consistency
of (e50) with respect to time reversal and space inversion should be accomplished. We address this task in section 6.2.
\subsection{Jordan and Mukunda results}
\subsubsection{For $\sigma(\underline P)=S_\mu^+$ and $\sigma(\underline P)=S_\mu^-$}
Jordan and Mukunda in \cite{JM} identified which operators $\bf Q$ are consistent with ({\it Q}.2.a), ({\it Q}.2.b) and (JM).
\par
In the case that corresponds to $\sigma(\underline P)=S_\mu^+$ and $U^+\mid_{{\mathcal P}_+^\uparrow}$ irreducible,
in the representations singled out in section 3 of the present article,
they find that for $s=0$  there is the unique solution ${\bf Q}={\bf F}$.
This result agrees with that obtained by us in section 5.2 where (JM) was not assumed, replaced with ({\it Q}.2.a).
\par
In general,
for $s\geq 0$ Jordan and Mukunda find that the solutions have the form
$$
{\bf Q}={\bf F}-\frac{a}{p_0(p_0+\mu)}({\bf P}\cdot{\bf S}){\bf P}+a{\bf S}-\frac{
{\bf P}\times{\bf S}}{\mu(p_0+\mu)},\hbox{ where }a\in\RR.\eqno(48)
$$
According to \cite{JM}, if $s>0$  such solution turns out to be non-commutative for any $a\in\RR$ , i.e. does not satisfy (Q.1).
Therefore, for $s>0$ the transformation property (JM) is inconsistent with the notion of position expressed by ({\it Q}.1), ({\it Q}.2.a), ({\it Q}.2.b).
\vskip.5pc
Identical results hold for $\sigma(\underline P)=S_\mu^-$.
\subsubsection{For $\sigma(\underline P)=S_\mu^+\cup S_\mu^-$}
In the case that corresponds to $\sigma(\underline P)=S_\mu^+\cup S_\mu^-$ and $U^+\mid_{{\mathcal P}_+^\uparrow}$ irreducible,
Jordan and Mukunda show that every solution for $\bf Q$, which satisfy (JM) and ({\it Q}.2.b) in our representation of section 3.2, must have the form
\vskip.5pc\noindent
${\bf Q}=\hat{\bf F}+\rho_1A(\sin B){\bf P}+\rho_2A(\cos B){\bf P}+$\hfill{(49)}\vskip.5pc
$\quad-\rho_1\left(\frac{\sin B}{p_0^2(p_0+\mu)}-\frac{2}{\mu}B'(\cos B)\right)({\bf p}\cdot{\bf S}){\bf P}
-\rho_2\left(\frac{\cos B}{p_0^2(p_0+\mu)}+\frac{2}{\mu}B'(\sin B)\right)({\bf p}\cdot{\bf S}){\bf P}+$
\vskip.5pc
$\quad+\frac{\rho_1}{p_0}(\sin B){\bf S}+\frac{\rho_1}{p_0}(\cos B){\bf S}
+\frac{{\bf P}\times {\bf S}}{p_0(p_0+\mu)}$,
\vskip.5pc\noindent
where $A=A({\bf p}^2)$, $B=B({\bf p}^2)$ are real functions of ${\bf p}^2$, and $B'=\frac{dB}{d({\bf p}^2)}({\bf p}^2)$,
$\rho_1=\left[\begin{matrix}0&1\cr 1&0\end{matrix}\right]$ and $\rho_2=\left[\begin{matrix}0&-i\cr i&0\end{matrix}\right]$
\par
We see that if $s=0$ then (49) coincides with the unique position operator, ${\bf Q}=\hat{\bf F}$, we have found in section 5.3, that is consistent with
({\it Q}.1) and ({\it Q}.2.a),({\it Q}.2.b).
\par
If $s>0$ the operator $\bf Q$ in (49) is not uniquely determined, so in general differs from $\hat{\bf F}$; but, in the case $s=\frac{1}{2}$, if $A=0$ and
$B\equiv$constant, i.e. $B'=0$,
then by a suitable unitary transformation that leaves $\bf P$, $P_0$, $\bf J$, $\bf K$ unaltered, (49) transforms into
$$
{\bf Q}=\hat{\bf F}+
\frac{\rho_2}{p_0^2(p_0+\mu)}({\bf p}\cdot{\bf S}){\bf P}
-\frac{\rho_2}{p_0}{\bf S}
+\frac{{\bf P}\times {\bf S}}{p_0(p_0+\mu)}\;.
\eqno(50)
$$
Jordan and Mukunda show that in this particular case
the theory is unitarily equivalent to the theory of Dirac for spin $\frac{1}{2}$ particles.
\subsection{Consistency of (JM) with $\TREV$ and $\SREV$}
In this section we check the consistency of the Jordan-Mukunda position operators with respect to time reversal and space inversion.
\subsubsection{Consistency in the case $\sigma(\underline P)=S_\mu^+$ and $\sigma(\underline P)=S_\mu^-$}
If the particle is characterized by the condition $\sigma(\underline P)=S_\mu^+$, with $U^+\mid_{{\mathcal P}_+^\uparrow}$ irreducible and $s=0$,
there is no problem of consistency of the Jordan and Mukunda transformation property (JM) with ({\it Q}.1), ({\it Q}.2.a) and ({\it Q}.2.b).
Indeed,
the unique operator ${\bf Q}={\bf F}$ determined in section 5.2 by ({\it Q}.1), ({\it Q}.2.a) and ({\it Q}.2.b)
coincides with the operator determined by Jordan and Mukunda, that satisfies (JM).
\vskip.5pc
In the case $s>0$, according to section 3.1.1, the anti-unitay time reversal operator is $\TREV=\tau{\mathcal K}\Upsilon$,
and the space inversion operator is $\SREV=\Upsilon$. By making use of these explicit operators and of (48), it turns out that
the condition $\SREV{\bf Q}=-{\bf Q}\SREV$ in ({\it Q}.2.a) implies $a=0$. Hence
$$
{\bf Q}={\bf F}-\frac{
{\bf P}\times{\bf S}}{\mu(p_0+\mu)}\;.\eqno(51)
$$
This operator satisfies $\TREV{\bf Q}={\bf Q}\TREV$.
However, according to Jordan and Mukunda, this operator is not commutative. Thus,
in this case there is no position operator that satisfies ({\it Q}.1), ({\it Q}.2.a) and ({\it Q}.2.b)
consistent with the transformation property (JM).
\vskip.5pc
The same result can be analogously obtained if $\sigma(\underline P)=S_\mu^-$.
\subsubsection{Consistency in the case $\sigma(\underline P)=S_\mu^+\cup S_\mu^-$}
The irreducible generalized projective representations with
$\sigma(\underline P)=S_\mu^+\cup S_\mu^-$ and irreducible $U^+\mid_{{\mathcal P}_+^\uparrow}$ are identified in section 3.2.
Apart from the case $\TREV$ anti-unitary and $\SREV$ unitary, implying $\sigma(\underline P)=S_\mu^+$ or $\sigma(\underline P)=S_\mu^-$,
all combinations $\TREV$, $\SREV$ concerning their unitary or anti-unitary character are possible.
\vskip.5pc
In section 5.3 we have seen that for $s=0$ there is a unique position operator ${\bf Q}=\hat{\bf F}$ that is consistent with ({\it Q}.1), ({\it Q}.2.a) and
({\it Q}.2.b). This coincides with the operator (49) that, according to Jordan and Mukunda, is consistent also with their transformation property (JM).
Therefore, for this particular case there is no consistency problem. However, according to section 5.3, this solution is valid only if $\TREV$ is unitary and $\SREV$ is {\sl anti-unitary}, namely $\TREV=\left[\begin{matrix}0&1\cr 1&0\end{matrix}\right]$ and $\SREV={\mathcal K}\left[\begin{matrix}0&1\cr 1&0\end{matrix}\right]$ or
$\SREV={\mathcal K}\left[\begin{matrix}0&1\cr -1&0\end{matrix}\right]$, contrary to the common conviction that $\TREV$ must be unitary and $\SREV$ also unitary
\cite{WeinBook}.
\vskip.5pc
Now we have to address the theories where $s>0$. We begin by checking the cases where $\TREV$ is anti-unitary. According to Prop. 3.2, $\TREV=\tau{\mathcal K}\Upsilon$; by making use of (49) we find that $\TREV{\bf Q}={\bf Q}\TREV$ holds if and only if $\sin B=0$ and $B'\cos B=0$. Hence $\bf Q$ becomes
$$
{\bf Q}=\hat{\bf F}+\rho_2A(\cos B){\bf P}
-\rho_2\frac{\cos B}{p_0^2(p_0+\mu)}({\bf p}\cdot{\bf S}){\bf P}
+\frac{\rho_2}{p_0}(\cos B){\bf S}
+\frac{{\bf P}\times {\bf S}}{p_0(p_0+\mu)}.\eqno(52)
$$
Being $\TREV$ anti-unitary, also $\SREV$ must be anti-unitary, i.e.  $\SREV=\left[\begin{matrix}0&\tau\cr \tau&0\end{matrix}\right]{\mathcal K}$
or $\SREV=\left[\begin{matrix}0&\tau\cr -\tau&0\end{matrix}\right]{\mathcal K}$.
If $\SREV=\left[\begin{matrix}0&\tau\cr -\tau&0\end{matrix}\right]{\mathcal K}$, then
$\SREV{\bf Q}=-{\bf Q}\SREV$ would imply $\frac{\partial}{\partial p_j}=0$, for all $j$.
Therefore this possibility must be excluded.
If $\SREV=\left[\begin{matrix}0&\tau\cr \tau&0\end{matrix}\right]{\mathcal K}$, by making use of (52) we find that
$\SREV{\bf Q}=-{\bf Q}\SREV$ holds if and only if $\cos B=0$, that is impossible because $\sin B=0$. Thus, for $s>0$ there is no
position operator $\bf Q$ that is consistent with ({\it Q}.1), ({\it Q}.2.a), ({\it Q}.2.b) and the transformation property (JM)
in a theory where $\TREV$ is anti-unitary.
\vskip.5pc
Let us check the case where $\TREV$ is unitary, i.e. $\TREV=\left[\begin{matrix}0&1\cr 1&0\end{matrix}\right]$. By making use of (49)
we find that $\TREV{\bf Q}={\bf Q}\TREV$ holds if and only if $\cos B=0$, so that $\bf Q$ becomes
$${\bf Q}=\hat{\bf F}+\rho_1A(\sin B){\bf P}-
\rho_1\frac{\sin B}{p_0^2(p_0+\mu)}({\bf p}\cdot{\bf S}){\bf P}\eqno(53)
$$
$$
-\frac{2\rho_2}{\mu}B'(\sin B)({\bf p}\cdot{\bf S}){\bf P}
+\frac{\rho_1}{p_0}(\sin B){\bf S}
+\frac{{\bf P}\times {\bf S}}{p_0(p_0+\mu)}.
$$
If $\SREV$ is unitary, i.e. $\SREV=\Upsilon$ or $\SREV=\left[\begin{matrix}1&0\cr 0&-1\end{matrix}\right]$. In the case $\SREV=\Upsilon$,
$\SREV{\bf Q}=-{\bf Q}\SREV$ implies $\sin B=0$, that is impossible, because $\cos B=0$.
In the case $\SREV=\left[\begin{matrix}1&0\cr 0&-1\end{matrix}\right]$, then $\SREV{\bf Q}=-{\bf Q}\SREV$ implies
$\rho_1A(\sin B){\bf P}+\frac{{\bf P}\times {\bf S}}{p_0(p_0+\mu)}=0$, that is false.
Thus, if both $\TREV$ and $\SREV$ are unitary, there is no position operator consistent with
({\it Q}.1),  ({\it Q}.2.a), ({\it Q}.2.b) and (JM).
\par
The last possibility is the case that $\SREV$ is anti-unitary, where
$\SREV=\left[\begin{matrix}0&\tau\cr \tau&0\end{matrix}\right]{\mathcal K}$
or $\SREV=\left[\begin{matrix}0&\tau\cr -\tau&0\end{matrix}\right]{\mathcal K}$.
If $\SREV=\left[\begin{matrix}0&\tau\cr -\tau&0\end{matrix}\right]{\mathcal K}$ then
$\SREV{\bf Q}=-{\bf Q}\SREV$ cannot hold because it would imply $\frac{\partial}{\partial p_j}=0$, for all $j$.
If $\SREV=\left[\begin{matrix}0&\tau\cr \tau&0\end{matrix}\right]{\mathcal K}$, then by making use of (53) we find that
$\SREV{\bf Q}=-{\bf Q}\SREV$ holds if and only if $A=0$. Now, $\sin B=\pm 1$, since $\cos B=0$, and (53) becomes
$$
{\bf Q}=\hat{\bf F}\mp
\rho_1\frac{\rho_1}{p_0^2(p_0+\mu)}({\bf p}\cdot{\bf S}){\bf P}
\pm\frac{\rho_1}{p_0}{\bf S}
+\frac{{\bf P}\times {\bf S}}{p_0(p_0+\mu)}.
\eqno(54)
$$
Thus, the unique quantum theory with $\sigma(\underline P)=S_\mu^+\cup S_\mu^-$, $U^+\mid_{{\mathcal P}_+^\uparrow}$ irreducible and $s>0$, where
the position operator is consistent with ({\it Q}.1),  ({\it Q}.2.a), ({\it Q}.2.b) and (JM), has $\TREV$ unitary and anti-unitary
$\SREV=\left[\begin{matrix}0&\tau\cr \tau&0\end{matrix}\right]{\mathcal K}$, and the position operator is given by (54).
By fixing $\sin B=-1$, we have
$$
{\bf Q}=\hat{\bf F}+
\rho_1\frac{\rho_1}{p_0^2(p_0+\mu)}({\bf p}\cdot{\bf S}){\bf P}
-\frac{\rho_1}{p_0}{\bf S}
+\frac{{\bf P}\times {\bf S}}{p_0(p_0+\mu)}.
\eqno(55)
$$
We see that the position operator is different from the position operator of Dirac's theory given by (50).
However, by transforming every operator $R$ into $WRW^{-1}$, where $W=e^{-i\frac{1}{2}\rho_3\frac{\pi}{2}}$,
with $\rho_3=\left[\begin{matrix}1&0\cr 0&-1\end{matrix}\right]$, it turns out that
all generators $P_j$, $P_0$, $J_k$, $K_j$ are left invariant, while $\bf Q$ of (55) is turned into that of (50).
Therefore, Dirac's theory is just the unique theory where ({\it Q}.1),  ({\it Q}.2.a), ({\it Q}.2.b) and (JM) hold,
but, contrary to the common conviction, $\SREV$ is anti-unitary.
\section{Conclusions}
In order to develop the relativistic quantum theories of single free particle, we have
pursued an approach based on group theoretical methods, whose methodological commitments prevents from the shortcomings yielded by canonical quantization.
In doing so we have obtained four inequivalent complete theories for spin 0 particles that are coherent and that are not affected
by the problems of Klein-Gordon theory.
\par
For the case of non zero spin, our approach is yet unable to determine complete theories, because of the present
inability to determine explicit quantum transformation properties of position with respect to boosts.
Such transformation properties can be determined only in the complete theories for spin 0 particles, where they can be directly calculated,
all involved operators being explicitly known; they turn out to coincide with the transformation properties (JM)
proposed by Jordan and Mukunda \cite{JM} also for the non zero spin case. Then we have checked the consistency of (JM) with the theories developed in our approach.
\par
As a result we found that Dirac theory is the unique theory for $s=\frac{1}{2}$ and $\sigma(\underline P)=S_\mu^+\cup S_\mu^-$
such that (JM) are satisfied. This could be taken as an argument supporting the general validity of (JM).
On the other hand, (JM) is inconsistent with the existence of localizable particle with $\sigma(\underline P)=S_\mu^+$
and $s>0$.
There is  a way that certainly would solve the dilemma:
to derive,
according to the methodological commitments of the present work, the explicit mathematical relation that expresses the transformation properties of position
with respect to boosts from sound physical principles, analogously to what done to obtain ({\it Q}.2a) and ({\it Q}.2b).
Let us denote these aimed mathematical relations by (KQ).
Whenever such a deduction is successful, one of the following two results will
be obtained.
\begin{itemize}
\item[Either]
(KQ) are equivalent to (JM);
in this case there would be precise consequences; for instance, one consequence will be that
Poincar\'e invariance is incompatible with the existence of particles with $\sigma(\underline P)=S_\mu^+$, or $\sigma(\underline P)=S_\mu^-$, and
$s>0$.
Another consequence will be that if $\sigma(\underline P)=S_\mu^+\cup S_\mu^-$ and $s=\frac{1}{2}$, then the unique
theory consistent with Poincar\'e invariance is Dirac theory.
\item[Or]
(KQ) are not equivalent to (JM) in some theories. Let us suppose that equivalence fails
for $\sigma(\underline P)=S_\mu^+\cup S_\mu^-$ and $s=\frac{1}{2}$. In this case we should investigate
which of the candidates for position operators identified by (36), (37) in section 5.3 are consistent with (KQ). If they exist
at all, we would have determined a theory {\sl alternative} to that of Dirac, implied by physical principles
without canonical quantization.
\end{itemize}
The arguments above enforce the importance of working for a determination of the relation (KQ), in order to realize
a real advancement in the understanding of relativistic quantum theory of a particle.
\vskip.5pc
The investigation reported in this paper is restricted to {\sl free} particles theories.
The natural next step will be to extend the approach to {\sl interacting} particle theory, by keeping the same methodological
commitments here followed.
Interesting ideas to this aim have been traced by L\'evy-Leblond in his attempt \cite{JMLL2} based on transformation properties;
they could be fruitful when applied to the present framework.
The realization of such a program would be an important advancement.
In particular, it should be useful to understand the meaning of the different theories
in terms of different physical properties, as charge and others, and hence of different particles, which can emerge only as differences in the way of interacting.
Furthermore, it would be possible to verify whether also the problems envisaged in connection with interaction, such as those implied by the transition to lower levels,
disappear in the new theory.

\end{document}